\documentclass[10pt,final,doublecolumn]{IEEEtran}
\usepackage{amsmath}
\usepackage{latexsym}
\usepackage{graphicx}
\usepackage{bbding}
\usepackage{indentfirst}
\usepackage{algorithm,algorithmic}
\usepackage{setspace}
\usepackage{float}
\usepackage{epstopdf}
\usepackage{amssymb}
\usepackage{amsfonts}
\usepackage{enumerate}
\usepackage{multicol}
\usepackage{color}
\usepackage{slashbox}
\usepackage{bm}
\usepackage{amssymb}
\usepackage{stfloats}

\IEEEoverridecommandlockouts
\allowdisplaybreaks[4]

\begin{document}
\title{Exploiting Simultaneous Low-Rank and Sparsity in Delay-Angular Domain for Millimeter-Wave/Terahertz Wideband Massive Access}
\author{Xiaodan Shao, Xiaoming Chen, Caijun Zhong, and Zhaoyang Zhang
\thanks{Xiaodan Shao ({\tt shaoxiaodan@zju.edu.cn}), Xiaoming Chen ({\tt chen\_xiaoming@zju.edu.cn}), Caijun Zhong ({\tt caijunzhong@zju.edu.cn}), and Zhaoyang Zhang ({\tt ning\_ming@zju.edu.cn}) are with the College of Information Science and Electronic Engineering, Zhejiang University, Hangzhou 310027, China. }}\maketitle

\begin{abstract}
Millimeter-wave (mmW)/Terahertz (THz) wideband communication employing a large-scale antenna array is a promising technique of the sixth-generation ($6$G) wireless network for realizing massive machine-type communications (mMTC). To reduce the access latency and the signaling overhead, we design a grant-free random access scheme based on joint active device detection and channel estimation (JADCE) for mmW/THz wideband massive access. In particular, by exploiting the simultaneously sparse and low-rank structure of mmW/THz channels with spreads in the delay-angular domain, we propose two multi-rank aware JADCE algorithms via applying the quotient geometry of product of complex rank-$L$ matrices with the number of clusters $L$. It is proved that the proposed algorithms require a smaller number of measurements than the currently known bounds on measurements of conventional simultaneously sparse and low-rank recovery algorithms. Statistical analysis also shows that the proposed algorithms can linearly converge to the ground truth with low computational complexity. Finally, extensive simulation results confirm the superiority of the proposed algorithms in terms of the accuracy of both activity detection and channel estimation.
\end{abstract}


\IEEEpeerreviewmaketitle
\section{Introduction}
The rapid advance of the Internet-of-Things (IoT) spurs an explosive growth in the number of devices. It is predicted that the number of IoT devices will reach hundreds of billions in 2030. In the context of massive IoT, conventional grant-based random access schemes lead to a high access latency and a prohibitive signaling overhead \cite{ICC,grant_based}. To this end, grant-free random access schemes have been widely regarded as a candidate technique of $6$G wireless networks, where the active devices transmit their data signals without a grant from the base station (BS) after sending pre-assigned pilot sequences. Hence, the key of grant-free random access is active device detection at the BS based on the received pilot sequences. However, due to a large number of devices but the limited radio resources for massive access, active device detection is a challenging problem.

Active device detection for grant-free random access is a typical compressed sensing (CS) problem since only a small fraction of devices are active at an arbitrary time slot. Consequently, some CS approaches have been applied to solve the problem of active device detection \cite{AMPO}-\cite{shi}. The authors in \cite{AMPO} proposed an approximate message passing (AMP) algorithm to jointly detect active devices and estimate corresponding channel state information (CSI) by exploiting the statistics of the wireless channel. It was proved that if the number of BS antennas was sufficiently large, the detection error asymptotically approached zero. Furthermore, a structured group sparsity estimation algorithm was developed in \cite{shi} to simultaneously detect the
active devices and estimate the corresponding channels. Note that
the aforementioned approaches in \cite{AMPO}-\cite{shi} perform activity detection based on instantaneous received signals, which require exceedingly long pilot sequences in the scenario of massive access. To tackle this problem, the authors in \cite{covar} proposed covariance-based algorithms for active device detection by using the covariance matrix of received signals. A common problem of previous works on active device detection is that they all considered a narrowband channel in sub-6 GHz band. In fact, with the increasing demands on enhanced mobile broadband (eMBB) and massive machine-type communications (mMTC), mmW and even THz bands are suggested to be adopted in $6$G wireless networks \cite{mmw1}.

Although mmW/THz can provide a broad vacant band, they also bring some challenging problems in activity detection and channel estimation for grant-free random access. First, mmW/THz signals are easily absorbed by rain and fog, and thus attenuate fast \cite{THz1}. In order to provide a wide coverage, mmW/THz communications have to employ a large-scale antenna array at the BS \cite{mmw2}. In this context, the received signal at the BS is high dimensional in both spatial and frequency domains. Therefore, activity detection has a prohibitively high computational complexity. Second, mmW/THz wideband massive access based on a large-scale antenna array requires long pilot sequences for accurate channel estimation, leading to extremely high resource consumption \cite{Pilot}.
It is worth noting that mmW/THz channels may exhibit angular and delay spreads \cite{delayspread}-\cite{park}. Such spreads give rise to
a useful low-rank structure of the channel that, along with the sparsity, can be simultaneously explored to reduce the overhead of the channel estimation \cite{path,single}. This is different from clustered channel model introduced in \cite{Alkhateeb}, where the channel matrix has a rank identical to its sparsity level. However, compared with the sparsity, the low-rank structure does not provide extra structural information. Very recently in \cite{single} and \cite{single1}, the sparse and low-rank properties of certain wireless channels were jointly exploited for efficient CSI estimation. Due to the fact that the estimation under rank constrained conditions is in general NP-hard, the authors in \cite{single} and \cite{single1} adopted the rank blind algorithms, where a nuclear norm was applied to relax the rank constraint. However, the nuclear norm-based convex relaxation approaches fail to well incorporate the fixed-rank matrices for sparse signal recovery due to the poor structures. To the best of the authors' knowledge, mmW/THz wideband massive access based on joint activity detection and channel estimation has not been investigated before.

To fully exploit the simultaneously sparse and low-rank structure of mmW/THz channels for wideband massive access in $6$G wireless networks, this paper aims to design a multi-rank aware recovery algorithm based on an important property of mmW/THz channels that the rank of the channel matrix is equal to the number of clusters, which  is far smaller than the sparsity level of the channel \cite{single}. The contributions of this paper are as follows:

\begin{enumerate}

\item We design an mmW/THz wideband massive access framework for $6$G wireless networks in the presence of delay and angular spreads based on the simultaneously sparse and low-rank structure of the channel matrix.

\item We propose two novel multi-rank aware sparse recovery algorithms for JADCE of mmW/THz wideband massive access based on the first-order search on product manifold. Theoretic analysis and simulation results confirm their lower computational complexity over baseline algorithms.

\item We analyze the sparse-block and low-rank restricted isometry property (SB$\&$L-RIP) and the convergence of the proposed algorithms, and prove that they require a smaller number of measurements than the known bounds on measurements of conventional algorithms. Hence, the proposed algorithms can shorten pilot sequences of JADCE.

\end{enumerate}

The rest of this paper is organized as follows. Section II gives a $6$G mmW/THz wideband grant-free random access framework. Then, Section III proposes two multi-rank aware recovery algorithms for JADCE. Next, Section IV analyzes the convergence of the proposed algorithms. Afterward, Section V provides extensive simulation results. Finally, Section VI concludes the paper.

\emph{Notations}: We use bold letters to denote matrices or vectors, non-bold letters to denote scalars, $\mathbf{I}_N$ to denote the identity matrix of order $N$, $\mathbb{C}^{A\times B}$ to denote the space of complex matrices of size $A\times B$, $|\cdot|$ to denote the absolute value,  $\left\|\cdot\right\|_F$ to denote Frobenius norm of a matrix, $\|\cdot\|$ to denote spectral norm of a matrix.
For a complex symbol/matrix, $(\cdot)^{'}$, $(\cdot)^H$, and $(\cdot)^T$ denote its complex conjugate, conjugate transpose, and transpose, respectively.
$\left \| \mathbf{A} \right \|_{l_1}=\sum _{i,j}\left | \mathbf{A}_{i,j}\right |$ denotes the $l_1$ norm of the matrix $\mathbf{A}$. $l_{0}$ and $\text{rank}(\cdot)$ are defined as the number of nonzero elements and the rank of a matrix respectively. $\left \lfloor a \right \rfloor$ is the downward rounding operation of a real number $a$. $\mathbf{A}_{(:,i)}$ and $\mathbf{A}_{(i,:)}$ represents the $j$th column and $j$th row of the matrix $\mathbf{A}$ respectively. $\|\mathbf{A}\|_{2,1}$ is the $l_{21}$ norm defined as the sum of the $l_2$ norm of the rows of a matrix $\mathbf{A}$. $\otimes$ denotes Kronecker product. $\text{vec}(\cdot)$ denotes the operation that stacks the columns of a matrix. $|\cdot|_c$ denotes the cardinality of a set. $[N]$ denotes the set $\{1,2,\cdots,N\}$. $\binom{N }{\Theta }$ denotes mathematical combination with scalars $\Theta \leq N$. $\mathrm{diag}(\mathbf{x})$ denotes a diagonal matrix with the diagonal entries specified by vector $\mathbf{x}$.

\section{System Model}
We consider a $6$G mmW/THz wideband wireless network, where a BS equipped with $M$ antennas serves a massive number of single-antenna IoT devices. The orthogonal frequency division multiplexing (OFDM) technique with $B$ subcarriers centered at the baseband frequencies $\{2\pi i/T_s\}_{i=0}^{B-1}$ is employed to combat the frequency-selective fading of mmW/THz wideband channels. Herein, $T_s$ is the useful OFDM symbol duration, which does not include the cyclic prefix, and the maximum delay spread of all devices' channels is not longer than $\gamma T_s$ with $\gamma \leq1$.
It is assumed that all devices are
synchronized with the BS in time and frequency.
In mMTC scenarios, only a fraction of IoT devices are active at any given time slot. In this context, a grant-free random access scheme is applied to jointly detect active devices and estimate their corresponding CSI.

In general, for a given detection accuracy, the required length of pilot sequences depends on the number of active devices. In order to reduce the pilot overhead of wideband massive grant-free random access, the BS partitions the uplink devices to groups of $N$ devices each, and each group is assigned an exclusive subset of subcarriers $\mathcal{B}_p \subseteq \{0,1,\cdots,B-1\}$ with the cardinality $B_p=\left|\mathcal{B}_p\right|_c\leq B$. The $K=\left|\mathcal{K}\right|_c$ active devices within a group transmit their predetermined pilot sequences on these subcarriers and one pilot symbol occupies a subcarrier, where the signal support $\mathcal{K}$ denotes the collection of active devices for an arbitrary group at a given time slot. If device grouping is not employed, long pilot sequences are needed to obtain accurate CSI of these devices.
In addition, to reduce the computational complexity, only the received signals from a subset of BS antennas $\mathcal{M}_p \in \{0,1,\cdots,M-1\}$ with cardinality $M_p$
are used for JADCE. Since the BS can perform JADCE within a group independently, we consider an arbitrary group in the rest of this paper. For convenience, we define $\chi_n$ as the activity indicator with ${\chi_n} = 1$ if the $n$th device is active, and ${\chi_n} = 0$ otherwise. Thus, the frequency-domain received signal $\mathbf{Y}\in \mathbb{C}^{M_p\times B_p}$ at the BS can be expressed as
\begin{align}\label{rece}
\mathbf{Y}
&=\sum_{n=1}^{N}\chi_n\mathbf{P}_M\mathbf{H}_n\mathbf{P}_T\text{diag}({\boldsymbol{\alpha} _n})+\mathbf{Z},
\end{align}
where $\mathbf{H}_n\in \mathbb{C}^{M\times B}$ is the channel matrix from the $n$th IoT device to the BS over $B$ subcarriers, $\mathbf{Z}$ denotes an additive white Gaussian noise (AWGN) matrix with independent and identically
distributed (i.i.d.) entries $\mathcal{CN}(0,{\sigma}^2)$. $\boldsymbol{\alpha}_n\in \mathbb{C}^{B_p}$ is the pilot sequence assigned to the $n$th device with zero mean and unit variance i.i.d. complex Gaussian distribution and is known to the BS in advance. Herein, we introduce the antenna sampling matrices $\mathbf{P}_M\in \{0,1\}^{M_p\times M}$ whose each row contains a single $1$ in the position corresponding to the selected antenna, and $0$
elsewhere, and the subcarrier sampling matrices $\mathbf{P}_T \in \{0,1\}^{B\times B_p}$ whose each column contains a single $1$ in the position corresponding to the selected subcarrier, and $0$ elsewhere.

Now, we give the selection strategy of BS antennas and OFDM subcarriers. On the one hand, for an arbitrary group of $N$ devices, the set of partial antennas $\mathcal{M}_p$ is randomly and uniformly selected from the set $\{0,1,\cdots,M-1\}$ with cardinality $M_p$, and the set $\mathcal{M}_p$ is the same for all groups. Hence, the received signal in the spatial domain has a low dimension, leading to a scalable computational and efficient JADCE algorithm. On the other hand, for an arbitrary group with $N$ devices, the set of partial subcarriers $\mathcal{B}_p$ is randomly and uniformly selected from the set $\{0,1,\cdots,B-1\}$ with cardinality $B_p$, and the set $\mathcal{B}_p$ is exclusive for each group. Under this design, the assignment of subcarriers to multiple groups is equivalent to the random and uniform partition of all subcarriers.


According to the theory of electromagnetic wave propagation, the mmW/THz channel of the device $n$ is a superposition of a small number of clusters, i.e., $L_n$, characterized by their delay $\tau_{n,l} \in [0, \gamma T_s]$ and angle of arrival (AoA) $\theta_{n,l} \in [-\pi/2,\pi/2]$ \cite{THz1,spread2,single}.
Then, the uplink channel from the device $n$ to the BS can be expressed as \cite{mmchan}
\begin{align}\label{chanvec}
{\bar{\mathbf{h}}_n}\left( \tau_n  \right) = \mathop \sum \limits_{l = 1}^{L_n} \left[\jmath_{n,l}{\bf{a}}\left( {{\theta _{n,l}}} \right)\delta \left( {{\tau _n} - {\tau _{n,l}}} \right)\right],
\end{align}
with the array response vector $\mathbf{a}(\cdot): [-\pi/2,\pi/2]\rightarrow \mathbb{C}^{M}$, which maps the angular domain to the spatial domain, and is defined as
$
\mathbf{a}(\theta_{n,l})=[1,e^{-j2\pi d\sin(\theta_{n,l})},\cdots ,e^{-j2\pi(M-1)d\sin(\theta_{n,l})}]^T
$. Herein, $d$ is the spatial separation of the uniform linear array (ULA), $\delta (\cdot)$ is the Dirac function, $\tau_n$ is the delay of device $n$, and $\jmath_{n,l}$ is the complex amplitude of the $l$th cluster of device $n$.

By adopting the OFDM technique, the uplink channel frequency response of the $b$-th subcarrier corresponding to the impulse response in \eqref{chanvec} is given by
$
{\bf{h}}_n\left( b \right) =\mathop \sum \limits_{l = 1}^{L_n} {\jmath_{n,l}}{\bf{a}}\left( {{\theta _{n,l}}} \right){e^{ - j\frac{{2\pi }}{T_s}{\tau _{n,l}}\left( {b - 1} \right)}}.
$
Hence, the wideband channel matrix is given by
\begin{align}\label{jointch}
\mathbf{H}_n=[{\bf{h}}_n\left( 1 \right), \cdots,{\bf{h}}_n\left( B \right)]=\sum \limits_{l = 1}^{L_n}\left[\jmath_{n,l} {\bf{a}}\left( {{\theta _{n,l}}} \right) \mathbf{b}(\tau_{n,l})^H\right],
\end{align}
with the vector $
\mathbf{b}(\tau_{n,l})=[1,e^{-j2\pi\tau_{n,l}/T_s},\cdots,e^{-j2\pi\tau_{n,l}(B-1)/T_s}]^T.
$

Due to the high resolution in the spatial domain by deploying a large-scale antenna array at the BS, the channel $\mathbf{H}_n$ exhibits angular spread \cite{spread2}. In other words, there exist more than one dominant coefficients caused by one cluster. We first consider a case where the angular spread over the AoA domain is a result of rays from a common delay. As a result, the channel $\mathbf{H}_n$ can be expressed as
\begin{align}\label{scho}
\mathbf{H}_n=\sum_{l=1}^{L_n} \left[\left( \sum_{j=1}^{J_n }\varsigma_{n,l,j}\mathbf{a}(\theta_{n,l}-\phi_{n,l,j})\right )\mathbf{b}(\tau_{n,l})^H\right],
\end{align}
where $\phi_{n,l,j}$ denotes the angle shift with respect to the mean angle $\theta_{n,l}$, $J_n$ represents the number of angular shifts, and $\varsigma_{n,l,j}$ denotes the complex gain of the $l$th cluster with the $j$th angle shift.
Similarly, due to the high resolution in the frequency domain by utilizing untra-wide mmW/THz band, the channel $\mathbf{H}_n$ also exhibits delay spread \cite{mmchan}. For instance, for two close delays, the channel $\mathbf{H}_n$ can be further expressed as
\begin{align}\label{schq}
\mathbf{H}_n&=\sum_{l=1}^{L_n}\left[\!\left( \!\sum_{j=1}^{J_n }\varsigma_{n,l,j}\mathbf{a}(\theta_{n,l}\!-\!\phi_{n,l,j})\right ) \mathbf{b}(\tau_{n,l}\!-\!\varphi _{n,l,1})^H\right.\nonumber\\
&\left.+\!\left(\! \sum_{j=1}^{J_n }\bar{\varsigma}_{n,l,j}\mathbf{a}(\theta_{n,l}\!-\!\phi_{n,l,j})\!\right )\!\mathbf{b}(\tau_{n,l}\!-\!\varphi _{n,l,2})^H\right]\!,\!\!\!\!
\end{align}
where $\varphi _{n,l,i}$ denotes the delay shift with respect to the mean delay $\tau_{n,l}$ of the $l$th cluster, and $J_n$ represents the number of angle shifts. Since the two terms $\left( \sum_{j=1}^{J_n }\varsigma_{n,l,j}\mathbf{a}(\theta_{n,l}-\phi_{n,l,j})\right )$ and $\left( \sum_{j=1}^{J_n }\bar{\varsigma}_{n,l,j}\mathbf{a}(\theta_{n,l}-\phi_{n,l,j})\right )$ are highly correlated, namely the complex gain ${\varsigma}_{n,l,j}$ and $\bar{\varsigma}_{n,l,j}$ associated with closely delays have similar power \cite{single}, expression \eqref{schq} can be further simplified as
\begin{align}\label{schq1}
\mathbf{H}_n\!=\!\sum_{l=1}^{L_n} \!\left[\!\left( \sum_{j=1}^{J_n }\varsigma_{n,l,j}\mathbf{a}(\theta_{n,l}-\phi_{n,l,j})\!\right)\!\! \!\left(\sum_{i=1}^{2}\mathbf{b}(\tau_{n,l}-\varphi _{n,l,i})^H\!\right)\!\right].
\end{align}
As a generalization, we adopt the following geometric channel model to characterize the mmW/THz channel for the device $n$
\begin{align}\label{sch}
\mathbf{H}_n\!&=\!\sum_{l=1}^{L_n}\!\left[ \!\left( \sum_{j=1}^{J_n }\varsigma_{n,l,j}\mathbf{a}(\theta_{n,l}-\phi_{n,l,j})\!\!\right)\!\left( \sum_{i=1}^{I_n}\xi _{n,l,i}\mathbf{b}(\tau_{n,l}\right.\right.\nonumber\\
&\left.\left.-\varphi _{n,l,i})^H\!\right)\!\right]\!,
\end{align}
where $\xi _{n,l,i}$ denotes the complex gain of the $l$th cluster with the $i$th delay shift and $I_n$ represents the number of delay shifts.

According to the delay-angular domain characteristics, the
mmW/THz channel $\mathbf{H}_n$ in (\ref{sch}) can be rewritten as
\begin{align}\label{sch1}
\mathbf{H}_n&=\sum_{l=1}^{L_n}\left[\mathbf{A}_{\theta}\boldsymbol{\varsigma}_{n,l}\boldsymbol{\xi}_{n,l}^T\mathbf{A}_{\tau}^H
\right]=\mathbf{A}_{\theta}\left(\sum_{l=1}^{L_n}\boldsymbol{\varsigma}_{n,l}\boldsymbol{\xi}_{n,l}^T\right)\mathbf{A}_{\tau}^H \nonumber\\
&\triangleq \mathbf{A}_{\theta}\widetilde{\mathbf{X}}_n\mathbf{A}_{\tau}^H,
\end{align}
where $\widetilde{\mathbf{X}}_n\in \mathbb{C}^{M_1 \times D}$ is the delay-angle representation channel. Considering the case of on-grid channel parameters, the steering matrices $\mathbf{A}_{\theta}$ and $\mathbf{A}_{\tau}$ are sampled versions from the interval $[-\pi/2, \pi/2]$ and the interval $[0, T_s]$, respectively. They can be represented as
$
\mathbf{A}_\theta \triangleq[\mathbf{a}(0), \mathbf{a}(1/M_1), \cdots, \mathbf{a}((M_1-1)/M_1)]\in\mathbb{C}^{M\times M_1}$ and
$\mathbf{A}_\tau \triangleq [\mathbf{b}(0),\mathbf{b}(T_s/B),\cdots,\mathbf{b}((D-1)T_s/B)]\\ \in\mathbb{C}^{B\times D}$, where $M_1\geq M$ is the number of samples, and $D=\lfloor{\gamma  B}\rfloor$ is the channel delay spread in samples with $\gamma \leq1$.
$\boldsymbol{\xi}_{n,l}\in \mathbb{C}^{D}$ and $\boldsymbol{\varsigma}_{n,l}\in \mathbb{C}^{M_1}$ are the virtual representation of delay and angular gains over the $l$th cluster, respectively.

In general, delay and angular spreads of mmW/THz channels are limited and take up a small part of the whole delay-angular domain. As a result, for the $n$th device, any vector in $\{\boldsymbol{\xi}_{n,l}, \boldsymbol{\varsigma}_{n,l}\}_{l=1}^{L_n}$ is sparse. Let us define $p_{\boldsymbol{\xi}_{n,l}}$ and $p_{\boldsymbol{\varsigma},n,l}$ as the number of non-zeros entries of $\boldsymbol{\xi}_{n,l}$ and $\boldsymbol{\varsigma}_{n,l}$, respectively. Then, $p_n$ is a quantity that measures the
maximum angular/delay spread among all scattering clusters $L_n$, which can be expressed as
$
p_n=\max_{l=1,2,\cdots,L_n}(p_{\boldsymbol{\varsigma},n,l},p_{\boldsymbol{\varsigma},n,l}).
$ Note that the true steering matrix parameters are assumed to lie on the discretized grids. In practice, the true
parameters may not lie on the discretized grid, which
is referred to as grid mismatch. In the presence of
grid mismatch, the vector in $\{\boldsymbol{\xi}_{n,l}, \boldsymbol{\varsigma}_{n,l}\}_{l=1}^{L_n}$ will not exactly contain at most $p_n$ non-zero entries. Instead, the
number of non-zero entries will become larger due to the power
leakage caused by grid mismatch. In this paper, we ignore the grid mismatch issue. Consequently, $\widetilde{\mathbf{X}}_n$ has at most $p_nL_n$ nonzero rows and $p_nL_n$ nonzero columns with $p_nL_n\ll\text{min}(D, M_1)$. $p_n$ can be much larger than the average number of clusters $L_n$ for the device $n$ \cite{delayspread, spread2}. Consequently, $\widetilde{\mathbf{X}}_n$ in \eqref{sch1} has a low-rank structure with $\text{rank}(\widetilde{\mathbf{X}}_n)=L_n\ll p_n^2L_n$. This is different from \cite{mmchan}, where the channel matrix has a rank identical to its sparsity level. In this case, the received signal in (\ref{rece}) can be rewritten as
\begin{align}\label{recey}
\mathbf{Y}
&\!=\!\sum_{n=1}^{N}\mathbf{P}_M\mathbf{A}_{\theta}\mathbf{X}_n\mathbf{A}_{\tau}^H\mathbf{P}_T\text{diag}({\boldsymbol{\alpha} _n})+\mathbf{Z}\!=\!\sum_{n=1}^{N}\mathbf{B}\mathbf{X}_n\mathbf{A}_n\!+\!\mathbf{Z}
\nonumber\\
&=\bar{\mathbf{A}}_\theta\mathbf{X}\bar{\mathbf{A}}_{\tau}^H+\mathbf{Z},
\end{align}
where $\mathbf{X}_n=\chi_n\widetilde{\mathbf{X}}_n\in \mathbb{C}^{M_1 \times D}$ is the device state matrix of device $n$, $\mathbf{B}=\mathbf{P}_M\mathbf{A}_{\theta}\in \mathbb{C}^{M_p\times M_1}$, $\mathbf{A}_n=\mathbf{A}_{\tau}^H\mathbf{P}_T\text{diag}({\boldsymbol{\alpha} _n})\in \mathbb{C}^{D\times B_p}$, $\bar{\mathbf{A}}_\tau=[\text{diag}({\boldsymbol{\alpha} _1})^H\mathbf{P}_T^T\mathbf{A}_{\tau},\cdots,\text{diag}({\boldsymbol{\alpha} _N})^H\mathbf{P}_T^T\mathbf{A}_{\tau}]\in \mathbb{C}^{B_p\times DN}$, $\bar{\mathbf{A}}_\theta=\mathbf{P}_M\mathbf{A}_{\theta}\in \mathbb{C}^{M_p\times M_1}$, and $\mathbf{X}=[\mathbf{X}_0,\mathbf{X}_1,\cdots,\mathbf{X}_{N-1}]\in \mathbb{C}^{M_1\times DN}$.

Due to a sporadic traffic pattern of IoT applications, $\mathbf{X}$ in (\ref{recey}) actually is $K$-block sparse. In other words, there are $K$ blocks or sub-matrices containing nonzero elements. As analyzed above, each block, e.g. $\mathbf{X}_n$, is also sparse. Such a sparsity across blocks and within each block is called as sparse-block sparsity. In addition, the matrix $\mathbf{X}$ is typically low-rank, namely $\text{rank}(\mathbf{X}) \leq \min\left \{ {M_1,KL_{\max}} \right \}$ due to low-rank of $\mathbf{X}_n$ and the denser sampling $M_1$ for the angle domain, where $L_{\max}$ is the maximum value of $L_n$ for $n=\{1,2,\cdots,N\}$. The recovery of the simultaneously sparse-block and low-rank signal $\mathbf{X}$ can be formulated as the following sparse-block lasso problem \cite{sgl}:
\begin{align}
\label{sgl}
\!\!\!\!&\!\!\!\!\hat{\mathbf{X}}=\mathop \text{arg~min}\limits_{\mathbf{X}}~\nu\left \| \mathbf{X} \right \|_{l_{0}}+\nu_1\text{rank} (\mathbf{X})+\nu_2\sum_{n=1}^{N}\left \| \mathbf{X}_{n} \right \|_F\nonumber\\
\!\!\!\!&\!\!\!\! \textrm{s.t.}~~~ \left \| \mathcal{A} (\mathbf{X})-\mathbf{y} \right \|_2\leq \epsilon,
\end{align}
where $\mathbf{y}=\text{vec}(\mathbf{Y})$ and the linear mapping $\mathcal{A}(\cdot): \mathbb{C}^{M_1\times DN}\rightarrow \mathbb{C}^{B_pM_p} $ obeys
$
\mathcal{A} (\mathbf{X})=\hat{\mathbf{A}}\mathbf{x}$
with $\hat{\mathbf{A}}=\bar{\mathbf{A}}_{\tau}^{'}\otimes\bar{\mathbf{A}}_\theta$ and
$\mathbf{x}=\text{vec}(\mathbf{X})$. Note that the Frobenius norm in the objective of \eqref{sgl} encourages sparsity across blocks, while the $l_{0}$ norm encourages sparsity within each
block. Herein, $\epsilon$ is an upper bound on the variance of the noise, and the parameter $\nu>0$, $\nu_1>0$ and $\nu_2>0$ are tunable regularization parameters. In what follows, we design a JADCE algorithm for mmW/THz wideband massive access via solving the above sparse-block lasso problem.

\section{Design of Multi-Rank Aware JADCE}
In this section, we first propose an alternative approximation for problem (\ref{sgl}), and then develop two multi-rank aware JADCE algorithms based on two first-order search methods on the product manifold, followed by computational complexity analysis and comparison.

\subsection{Simultaneously Sparse-Block and Low-Rank Approximation}
It is difficult to solve the sparse-block and low-rank problem in (\ref{sgl}) directly due to the combination of $l_{0}$ norm, Frobenius norm, and low rank. Moreover, three regularization coefficients need to be designed in (\ref{sgl}), which brings more calculation burden to the signal recovery. To resolve these challenges, we propose to recover the simultaneously sparse-block and low-rank device state matrix $\mathbf{X}$ by combining $l_{0}$ norm and low rank under approximation conditions, instead of jointly using sparse-block lasso and low rank mentioned in (\ref{sgl}). According to the characteristics of mmW/THz channels, we start the approximation with the following definition.

{\emph{Definition 1}}: A matrix $\mathbf{X}\in \mathbb{C}^{M_1\times DN}$ is $u$-sparse block if
\begin{align}
\label{jsg}
\left \| \mathbf{X} \right \|_{\mathrm{sg}}=\sum_{n \in [N]}\Xi_np_n^2L_n\leq u,
\end{align}
with
\begin{align}
\label{jsgs}
\Xi_n=\left\{\begin{matrix}
1,~ \mathrm{if} ~\mathbf{X}_{\Upsilon_n}\neq\mathbf{0}, \\
0,~\mathrm{if} ~\mathbf{X}_{\Upsilon_n}=\mathbf{0},
\end{matrix}\right.
\end{align}
holds, where $\Upsilon_n$ with $|\Upsilon_n|_c=D, n=1,\cdots, N$, are disjoint subsets of the column index set $\{1,2,\cdots,DN\}$ of $\mathbf{X}$. Matrix $\mathbf{X}_{\Upsilon_n}\in \mathbb{C}^{M_1 \times D}$ denotes the projection of $\mathbf{X}$ onto the column set $\Upsilon_n$, which has at most $p_nL_n$ nonzero rows and $p_nL_n$ nonzero columns.

Then, based on the fact that the device state matrix $\mathbf{X}$ is simultaneously $u$-sparse block and low rank, we can verify that jointly imposing $l_{0}$ norm and low-rank constraint on the matrix $\mathbf{X}$ makes
\begin{align}\label{sgrank}
  \!\!\!\!&\!\!\!\!\hat{\mathbf{X}}=\mathop \text{arg~min}\limits_{\mathbf{X}}\nu\left \| \mathbf{X} \right \|_{l_{0}}+\nu_1\text{rank}(\mathbf{X})\nonumber\\
 \!\!\!\!&\!\!\!\! \textrm{s.t.}~~~ \left \| \mathcal{A} (\mathbf{X})-\mathbf{y} \right \|_2\leq \epsilon,
\end{align}
be a viable alternative to jointly imposing Frobenius norm, $l_{0}$ norm, and low-rank constraints as shown in the problem (\ref{sgl}).

To verify it, we start by introducing a specific restricted isometry property (RIP), called sparse-block and low-rank RIP (SB$\&$L-RIP) defined as follows.

{\emph{Definition 2}}: For positive integers $u$ and $r$, the linear map $\mathcal{A}(\cdot)$ satisfies the SB$\&$L-RIP, if for all simultaneous $u$-sparse block and rank $r$ matrices $\mathbf{X}$, it is true that
\begin{align}\label{sbl}
(1-\delta_{u,r})\left \| \mathbf{X} \right \|_F\leq \left \| \mathcal{A}(\mathbf{X}) \right \|_2\leq
(1+\delta_{u,r})\left \| \mathbf{X} \right \|_F,
\end{align}
where $\delta_{u,r}$ is the smallest constant for which the above property holds.

Compared to the traditional definitions of Rank-RIP \cite{rankrip} and Block-RIP \cite{blockrip}, the defined sparse-block and low-rank RIP provides a more restrictive property which holds for the intersection set of the low-rank and the $u$-sparse block matrices. In what follows,
Theorem 1 gives a bound on the required number of measurements that
suffice to ensure that $\mathcal{A}(\cdot)$ satisfies the SB$\&$L-RIP, and then the problem (\ref{sgrank}) with the map $\mathcal{A}(\cdot)$ can achieve a robust sparse-block and low-rank recovery.

\emph{Theorem 1}: Let the linear map $\mathcal{A}(\cdot): \mathbb{C}^{M_1\times DN}\rightarrow \mathbb{C}^{B_pM_p} $ obey the following condition for any $\mathbf{X}\in \mathbb{C}^{M_1\times DN}$, $0<\delta<1$, and $t>1$
\begin{align}\label{concen}
P_r\left(\left|\left \| \mathcal{A}(\mathbf{X}) \right \|_2^2-\left \| \mathbf{X} \right \|_F^2\right|> \delta\left \| \mathbf{X} \right \|_F^2\right)\leq \text{exp}(-cB_pM_p),
\end{align}
where $c$ is a fixed parameter for a given $t$. Given integers $u$ and $N$, the map $\mathcal{A}(\cdot)$ satisfies SB$\&$L-RIP of order $\bar{u}:=[1+(t-1)p_{\max}^2L_{\max}]u$ with a constant $\delta_{\bar{u},r}<\delta$ with a probability $\geq 1-\bar{C}e^{-\kappa_0}$, if the number of measurements fulfills the following condition:
\begin{align}
\label{measure}
\!\!\!B_pM_p\!&\geq\!\kappa_1 \!\left(\!\Theta \log\frac{N}{\Theta }\!+\!\Theta \!+\!\Theta p_{\max}L_{\max}\log\frac{D}{p_{\max}L_{\max}}\!\right.\nonumber\\
&\left.+\Theta p_{\max}L_{\max}\!+\!\left(\Theta p_{\max}L_{\max}\!+\!M_1\!+\!1\right)r\!\right)\!,
\end{align}
where $\kappa_0$, $\kappa_1$, and $\bar{C}$ are constants for a given $\delta_{\bar{u},r}$, $\Theta =\bar{u}/p_{\min}^2L_{\min}$, $p_{\max}$ and $p_{\min}$ are the maximum and the minimum value of $p_n$ for $n=\{1,2,\cdots,N\}$, respectively, and $L_{\max}$ and $L_{\min}$ are the maximum and the minimum value of $L_n$ for $n=\{1,2,\cdots,N\}$, respectively.

\begin{IEEEproof}
Please refer to Appendix A.
\end{IEEEproof}

According to Theorem 1, the problem (\ref{sgrank}) with a map $\mathcal{A}(\cdot)$ ($\mathcal{A}(\mathbf{X}) \neq \mathbf{0}$) achieves a robust sparse-block and low-rank recovery of order $\bar{u}$ with a probability greater than $ 1-\bar{C}e^{-\kappa_0}$. Thus, we can remove the Frobenius norm constraint in problem \eqref{sgl}. Although there is no term to promote sparsity across blocks in (13), the $l_0$ norm promoting sparsity of $\mathbf{X}$ is enough for recovering $\mathbf{X}$ accurately. Noting that removing the Frobenius norm constraint in \eqref{sgl} does not change the search space of $\mathbf{X}$, since the both problems \eqref{sgl} and \eqref{sgrank} search the solution among the whole complex region.

\emph{Remark 1}: Note that $\Theta =\bar{u}/p_{\min}^2L_{\min}$ is equivalent to the number of active devices. Theorem 1 gives the relationship among the length of pilot sequences, the number of BS antennas, and the number of active devices for achieving reliable detection. In the case of a small $ut$ (a reasonable assumption in mmW/THz wideband massive access systems), the required number of measurements derived from Theorem 1 for the recovery of simultaneously $u$-sparse block and low-rank matrices are significantly smaller than currently known measurements bounds, which are derived for simultaneously sparse and low-rank matrices recovery \cite{trad2015}. See Appendix B for detailed proof.

\subsection{Multi-Rank Aware Pursuit}
As analyzed in Section II, the rank of the device state matrix $\mathbf{X}_n$ is not random but equals to the number of clusters $L_n$. Notice that the number of clusters of mmW/THz channels is usually very limited, e.g. 2-3, which can be measured by channel tracking \cite{path}. Inspired by it, we seek an approximated solution to (\ref{sgrank}) via the following multi-rank aware problem
\begin{align}\label{rankawa}
\!\!\!\!\!\!&\!\!\!\!\!\!\mathop \text{arg~min}\limits_{\left\{\mathbf{X}_n\right \}_{n=1}^N}\frac{1}{2}\left \| \sum_{n=1}^{N}\mathbf{B}\mathbf{X}_n\mathbf{A}_n-\mathbf{Y} \right \|_F^2 +\nu\sum_{n=1}^{N}\left \| \mathbf{X}_n \right \|_{l_{1}}\nonumber\\
\!\!\!\!\!\!&\!\!\!\!\!\!\textrm{s.t.}~~~ \text{rank}(\mathbf{X}_n)=L_{\max},~~~n=1,2,\cdots,N
\end{align}
Herein, we relax the rank of inactive device state matrices to $L_{\max}$, which is the maximum number of clusters, i.e., $L_{n}$, $n=\{1,2,\cdots,N\}$. Since device state matrices of inactive devices infinitely approach to zero matrices, even though we solve a relaxed problem, the rank relaxation would not change the solution to the original problem (\ref{sgrank}). Moreover, due to non-convexity of $l_{0}$ norm, it is quite natural to relax $l_{0}$ norm with $l_{1}$ norm. Because of multi-rank constraints, the problem (\ref{rankawa}) is still nonconvex and NP-hard. To tackle this challenge, we exploit the quotient manifold geometry of the product of rank-$L_{\max}$ matrices.

In general, the algorithms based on the quotient manifold are intractable for non-square fixed-rank matrix problem \eqref{rankawa} in the complex field. For such a problem, an efficient approach is the Burer-Monteiro factorization \cite{bur}, which optimizes over the low-rank factors of a matrix, but not the full matrix. The authors in \cite{yang}
have used Burer-Monteiro lifting to solve the topological cooperation problem. Inspired by these, we apply Burer-Monteiro factorization and semidefinite lifting to the multiple fixed-rank optimization problem (\ref{rankawa}). Specifically, we first carry out the low-rank factorization as $\mathbf{X}_n=\mathbf{J}_n\mathbf{R}_n^H$, where $\mathbf{J}_n\in \mathbb{C}^{M_1\times L_{\max}}$ and $\mathbf{R}_n \in \mathbb{C}^{D\times L_{\max}}$ are full column-rank matrices, whose columns are
linearly independent. Then, by defining $\mathbf{S}_n=\left[ \mathbf{J}_n^H,
\mathbf{R}_n^H \right]^H \in \mathbb{C}^{(D+M_1)\times L_{\max}}$, we can lift $\mathbf{X}_n$ in a factored form as follows:
\begin{equation}\label{FAC}
\mathbf{S}_n\mathbf{S}_n^H=\left[ \begin{array}{l}
\mathbf{J}_n\mathbf{J}_n^H ~~~\mathbf{J}_n\mathbf{R}_n^H\\
\mathbf{R}_n\mathbf{J}_n^H~~~\mathbf{R}_n\mathbf{R}_n^H
\end{array} \right].
\end{equation}
Next, two auxiliary matrices $
\mathbf{P}_1=\left [ \mathbf{I}_{D}~~\mathbf{0} \right ]\in\mathbb{C}^{M_1\times (D+M_1)}$ and $\mathbf{P}_2=\left[ \mathbf{0}~~
\mathbf{I}_{M_1} \right]^T\in\mathbb{C}^{ (D+M_1)\times D}$ are introduced, where $\mathbf{I}_{D}$ and $\mathbf{I}_{M_1}$ denote the identity matrices of order $D$ and $M_1$, respectively.
Upon multiplying both sides of $\mathbf{S}_n\mathbf{S}_n^H$ by
$\mathbf{P}_1$ and $\mathbf{P}_2$, $\mathbf{X}_n$ satisfies the factorization $\mathbf{X}_n=\mathbf{P}_1\mathbf{S}_n\mathbf{S}_n^H\mathbf{P}_2$.

Furthermore, we define the product manifold $\mathcal{M}^N$ as
\begin{align}\label{produ}
&\mathcal{M}^N=\mathcal{M}_1 \times \mathcal{M}_2\times\cdots \times\mathcal{M}_N,\nonumber\\
&\mathcal{M}_n=\left \{ \mathbf{S}_n\in \mathbb{C}^{(D+M_1)\times L_{\max}}:\text{rank}(\mathbf{S}_n)=L_{\max} \right \},
\end{align}
where $\mathcal{M}^N$ is the set of matrices $\left \{ \mathbf{S}_n \right \}_{n=1}^N$, and $\mathcal{M}_n$ is a non-compact Stiefel manifold denoting a set of all $(D + M_1) \times L_{\max}$ matrices whose columns are linearly independent. Consequently, the problem (\ref{rankawa}) can be recast as the following unconstrained problem with the full column rank optimization variables
$\mathbf{S}_n\in \mathbb{C}^{(D+M_1)\times L_{\max}}$:
\begin{align}\label{lift}
\mathop \text{arg~min}\limits_{\left \{ \mathbf{S}_n \right \}_{n=1}^N\in \mathcal{M}^N} &f(\left \{ \mathbf{S}_n\right \}_{n=1}^N)\!=\!\frac{1}{2}\left \| \sum_{n=1}^{N}\mathbf{B}\mathbf{P}_1\mathbf{S}_n\mathbf{S}_n^H\mathbf{P}_2\mathbf{A}_n-\mathbf{Y} \right \|_F^2\nonumber\\
&+\nu\sum_{n=1}^{N}\sum_{i,j}\left(\left |\mathbf{v}_i\mathbf{P}_1\mathbf{S}_n\mathbf{S}_n^H\mathbf{P}_2\mathbf{v}_j  \right |\right.\nonumber \\
&\left.-\frac{1}{\varrho }
\ln\left(1+\varrho \left |\mathbf{v}_i\mathbf{P}_1\mathbf{S}_n\mathbf{S}_n^H\mathbf{P}_2\mathbf{v}_j  \right |\right)\right),
\end{align}
where $\varrho >0$ is a tunable parameter, and $\mathbf{v}_i$ and $\mathbf{v}_j$ are used to extract the element located in the $i$th row and the $j$th column of the matrix $\mathbf{P}_1\mathbf{S}_n\mathbf{S}_n^H\mathbf{P}_2$. Notice that $l_{1}$ norm in problem (\ref{rankawa}) is nonsmooth, which hinders to solve the problem. We therefore have replaced the element of $ \left \| \mathbf{X}_n \right \|_{l_1}$ with the logarithmic smoothing function $g(x)=\left | x \right |-\frac{1}{\varrho}\ln(1+\varrho\left | x \right |)$, which is differentiable over $x$ at $0$ \cite{Pilot}. After the solution $\hat{\mathbf{S}}_n,\forall n$ of problem (\ref{lift}) is derived, the original solution $\hat{\mathbf{X}}_n, \forall n$ can be obtained by letting $\hat{\mathbf{X}}_n=\mathbf{P}_1\hat{\mathbf{S}}_n\hat{\mathbf{S}}_n^H\mathbf{P}_2, \forall n$.

\subsection{First-Order Search On Product Manifold}
Although the problem (\ref{lift}) is smooth, the factorization $\mathbf{S}_n\mathbf{S}_n^H$ for rank-$L_{\max}$ matrices is non-unique, which prevents us finding the optimal solution. Therefore, we develop a set of equivalence classes encoding the invariance map on product manifold $\mathcal{M}^N$ with $n=1, 2, \cdots, N$ in an abstract search space of the form
\begin{align}
[\boldsymbol{\mathcal{S}}]=\left \{ [\mathbf{S}_n] \right \}_{n=1}^N=&\left \{ \mathbf{S}_n\mathbf{Q}_n: \mathbf{Q}_n^H\mathbf{Q}_n=\mathbf{Q}_n\mathbf{Q}_n^H=\mathbf{I}\right., \nonumber\\
 &\left.\mathbf{Q}_n\in \mathbb{C}^{L_{\max}\times L_{\max}} \right\}_{n=1}^N.
\end{align}
Herein, $[\boldsymbol{\mathcal{S}}]$ is called as the quotient space denoted by $\mathcal{M}^N/\sim $, where $\mathcal{M}^N$, as the product of non-compact Stiefel manifold $\mathcal{M}_n$, is regarded as the total space. Consequently, the problem (\ref{lift}) can be transformed as
\begin{align}\label{lift1}
\mathop \text{arg~min}\limits_{\left \{ [\mathbf{S}_n] \right \}_{n=1}^N\in \mathcal{M}^N/\sim} f(\left \{ [\mathbf{S}_n]\right \}_{n=1}^N).
\end{align}
Because the manifold topology of the product manifold is equivalent to the product topology \cite{pa}, the problem \eqref{lift1} on the product manifold $\mathcal{M}^N$ can be processed on individual manifold $\mathcal{M}_n$, and the tangent space to $\mathcal{M}^N$ at $\boldsymbol{\mathcal{S}}$ given by $\mathcal{T}_{\boldsymbol{\mathcal{S}}}\mathcal{M}^N$ can be viewed as the product of the tangent spaces to $\mathcal{M}_n$ at $\mathbf{S}_n$ given by $\mathcal{T}_{\mathbf{S}_n}\mathcal{M}
$ for $n=1, 2, \cdots, N$.

In the context of individual manifold, we provide a Riemannian metric \cite{pa}, which is the smoothly varying inner product, namely
\begin{align}\label{inner}
   g_{\mathbf{S}_n}(\boldsymbol{\xi}_{\mathbf{S}_n},\boldsymbol{\eta}_{\mathbf{S}_n})
=\text{Tr}(\boldsymbol{\xi}_{\mathbf{S}_n}^H\boldsymbol{\eta}_{\mathbf{S}_n}+\boldsymbol{\eta}_{\mathbf{S}_n}^H\boldsymbol{\xi}_{\mathbf{S}_n}),
\boldsymbol{\xi}_{\mathbf{S}_n},\boldsymbol{\eta}_{\mathbf{S}_n} \in \mathcal{T}_{\mathbf{S}_n}{\mathcal{M}_n}.
\end{align}
The individual projection of any direction $\boldsymbol{\xi}_{\mathbf{S}_n}$ onto the horizontal space $\mathcal{H}_{\mathbf{S}_n}$ at $\mathbf{S}_n$ is given by $\Pi_{\mathcal{H}_{\mathbf{S}_n}}(\boldsymbol{\xi}_{\mathbf{S}_n})=\boldsymbol{\xi}_{\mathbf{S}_n}-\mathbf{S}_n\mathcal{B}$,
where $\mathcal{B}$ is a complex matrix of size $L_{\max} \times L_{\max}$, which is the solution of the following Lyapunov equation: $
\mathbf{S}_n^H\mathbf{S}_n\mathcal{B}+\mathcal{B}\mathbf{S}_n^H\mathbf{S}_n=\mathbf{S}_n^H\boldsymbol{\xi}_{\mathbf{S}_n}-\boldsymbol{\xi}_{\mathbf{S}_n}^H\mathbf{S}_n$ \cite{pa}.

Given the Riemannian metric in (\ref{inner}), it is likely to deduce the Riemannian gradient on manifolds represented in the tangent space for minimizing the objective function in (\ref{lift1}). Define $\mathbf{S}=\left [ \mathbf{S}_1^H,\mathbf{S}_2^H,\cdots,\mathbf{S}_N^H \right ]^H$, and then the Riemannian gradient $\text{grad}_{\mathbf{S}_n}f$ is the unique operator satisfying
$
  g_{\mathbf{S}_n}(\text{grad}_{\mathbf{S}_n}f,\boldsymbol{\eta}_{\mathbf{S}_n})
  =\mathcal{D}_{\mathbf{S}_n}f(\mathbf{S})[\boldsymbol{\eta}_{\mathbf{S}_n}],\forall \boldsymbol{\eta}_{\mathbf{S}_n}\in \mathcal{T}_{\mathbf{S}_n}\mathcal{M}
$,
where
$
\mathcal{D}_{\mathbf{S}_n}f(\mathbf{S})[\boldsymbol{\eta}_{\mathbf{S}_n}]:=\lim_{t\rightarrow 0}\frac{f(\mathbf{S})|_{\mathbf{S}_n+t\boldsymbol{\eta}_{\mathbf{S}_n}}-f(\mathbf{S})|_{\mathbf{S}_n}}{t}
$
is the directional derivative of $f(\mathbf{S})$ with respect to $\mathbf{S}_n$ along the direction $\boldsymbol{\eta}_{\mathbf{S}_n}$.
Thus, the Riemannian gradient, i.e., the horizontal lift of $\mathcal{D}_{\mathbf{S}_n}f(\mathbf{S})$ to $\mathcal{T}_{\mathbf{S}_n}{\mathcal{M}_n}$, is given by
$
\text{grad}_{\mathbf{S}_n}f=\Pi_{\mathcal{H}_{\mathbf{S}_n}}\left( \frac{1}{2} \mathcal{D}_{\mathbf{S}_n}f(\mathbf{S})\right),
$ where $\mathcal{D}_{\mathbf{S}_n}f(\mathbf{S})$ is the Euclidean gradient of $f(\mathbf{S})$ with respect to $\mathbf{S}_n$.
By substituting the objective function of (\ref{lift}) into $\mathcal{D}_{\mathbf{S}_n}f(\mathbf{S})$, we can get
\begin{align}\label{grad1}
&\mathcal{D}_{\mathbf{S}_n}\!f(\mathbf{S})=
\frac{1}{2}\mathbf{P}_1^T\mathbf{B}^H\!\!\left(\!\sum_{n=1}^{N}\mathbf{B}\mathbf{P}_1\mathbf{S}_n\mathbf{S}_n^H\mathbf{P}_2\mathbf{A}_n\!-\!\mathbf{Y}\!\!\right)\!\!\mathbf{A}_n^H\mathbf{P}_2^T \mathbf{S}_n\nonumber \\
&+\frac{1}{2}\mathbf{P}_2\mathbf{A}_n\!\left(\!\sum_{n=1}^{N} \mathbf{B}\mathbf{P}_1\mathbf{S}_n\mathbf{S}_n^H\mathbf{P}_2\mathbf{A}_n\!-\!\mathbf{Y}\!\!\right)^H\!\mathbf{B}\mathbf{P}_1\mathbf{S}_n \nonumber\\
&\!+\!\mathbf{P}_1^T\!\left(\!(\nu\varrho\mathbf{P}_1\mathbf{S}_n\mathbf{S}_n^H\mathbf{P}_2)\!./\! (\mathbf{I}\!+\!\text{sgn}(\mathbf{P}_1\mathbf{S}_n\mathbf{S}_n^H\mathbf{P}_2)\varrho)\!\right)\!\mathbf{P}_2^T\mathbf{S}_n\nonumber\\
&+\mathbf{P}_2  \left ( (\ \nu\varrho\mathbf{P}_1\mathbf{S}_n\mathbf{S}_n^H\mathbf{P}_2)\!./\!  (\mathbf{I}\!+\!\text{sgn}(\mathbf{P}_1\mathbf{S}_n\mathbf{S}_n^H\mathbf{P}_2)\varrho) \right )^H\mathbf{P}_1\mathbf{S}_n,
\end{align}
where $./$ denotes the element-wise division, and the $(m,d)$th element of the matrix $\text{sgn}(\mathbf{X})$ equals to $1$ if $x_{m,d}> 0$, equals to $0$ if $x_{m,d}= 0$, and equals to $-1$ if $x_{m,d}< 0$.
Herein, $x_{m,d}$ denotes the $(m,d)$th element of the matrix $\mathbf{X}$.

We now need to determine the search direction in the tangent space $\mathcal{T}_{\mathbf{S}_n}{\mathcal{M}_n}$. A retraction on a manifold $\mathcal{M}_n$ is a smooth
mapping from the tangent bundle $\mathcal{T}_{\mathbf{S}_n}{\mathcal{M}_n}$ onto $\mathcal{M}_n$.
It can be used to locally transform an optimization problem on the manifold $\mathcal{M}_n$ into an optimization problem on the more friendly vector space $\mathcal{T}_{\mathbf{S}_n}{\mathcal{M}_n}$  \cite{pa}. In this paper, the following efficient retraction is adopted for ensuring that each update of the search variable is located on the manifold
\begin{align}\label{sta}
\mathcal{R}_{\mathbf{S}_n}(\mu\boldsymbol{\eta}_{{\mathbf{S}}_n})
    =\mathbf{S}_n+\mu\boldsymbol{\eta}_{{\mathbf{S}}_n},
\end{align}
where $\mu>0$ is the step size and $\boldsymbol{\eta}_{{\mathbf{S}}_n}\in \mathcal{T}_{\mathbf{S}_n}{\mathcal{M}_n}$ is a search direction. Such a retraction can provide a computationally efficient approach to smoothly select a moving curve on a manifold. Now, we provide two low-complexity first-order search methods to determine the direction $\boldsymbol{\eta}_{{\mathbf{S}}_n}$.

The first method sets the negative weighted Riemannian gradient in the $t$th iteration as the search direction, which is given by $
\boldsymbol{\eta}_{{\mathbf{S}}_n}^t=-\frac{1}{ g_{\mathbf{S}_n^t}(\mathbf{S}_n^t,\mathbf{S}_n^t)}\text{grad}_{\mathbf{S}_n^t}f.
$
We call the recovery algorithm based on the negative weighted Riemannian gradient as a Riemannian gradient decent-based multi-rank aware sparse recovery (RG-MRAS) algorithm.

The second method chooses the search direction as
\begin{align}\label{sear}
\boldsymbol{\eta}_{{\mathbf{S}}_n}^{t+1}=-\text{grad}_{\mathbf{S}_n}f+o_n^t\boldsymbol{\Im}_{\mu^t{\boldsymbol{\eta}_{{\mathbf{S}}_n}^t}}({\boldsymbol{\eta}_{{\mathbf{S}}_n}^t}),
\end{align}
where $\boldsymbol{\Im}_{\mu{\boldsymbol{\eta}_{{\mathbf{S}}_n}}}({\boldsymbol{\eta}_{{\mathbf{S}}_n}})$
is the vector transport operator so that $\boldsymbol{\eta}_{{\mathbf{S}}_n}$ is
transported from $\mathcal{T}_{\mathbf{S}_n}\mathcal{M}
$ to $\mathcal{R}_{\mathbf{S}_n}(\mu\boldsymbol{\eta}_{{\mathbf{S}}_n})$.
The vector transport is collinear with ${\boldsymbol{\eta}_{{\mathbf{S}}_n}^t}$ \cite{pa}.
The parameter $o_n^t$ in the Polak-Ribiere form is selected as below:
\begin{align}\label{sear1}
o_n^t=\frac{g_{\mathbf{S}_n^{t}}\left(\text{grad}_{\mathbf{S}_n^{t+1}}f,\text{grad}_{\mathbf{S}_n^{t+1}}f-\boldsymbol{\Im}_{\mu^t{\boldsymbol{\eta}_{{\mathbf{S}}_n}^t}}
(\text{grad}_{\mathbf{S}_n^t}f)\right)}{g_{\mathbf{S}_n^t}\left(\text{grad}_{\mathbf{S}_n^t}f,\text{grad}_{\mathbf{S}_n^t}f\right)}.
\end{align}
Correspondingly, we call the algorithm based on the above Riemannian conjugate gradient as a Riemannian conjugate gradient-based multi-rank aware sparse recovery (RC-MRAS) algorithm.

A key step to ensure a meaningful convergence for the search of $\mathbf{S}_n$ is to initialize the proposed search methods RG-MRAS and RC-MRAS with a certain point inside the basin of attraction.
Herein, the basin of attraction of the attracting fixed point $\mathbf{S}_n^*$ is the set of all matrices $\mathbf{S}_n$ such that for a given initial value $\mathbf{S}_n^0=\mathbf{S}_n$, the variable in the $t$th iteration $\mathbf{S}_n^t$ asymptotically approaches $\mathbf{S}_n^*$ as $t\rightarrow \infty$. Traditionally, the spectral method can provide an appealing initialization. However, since the dimension of the term $\mathbf{B}^H\mathbf{Y}\mathbf{A}_n^H$ is high in the spectral method, obtaining its leading eigenvector has a prohibitively computational complexity. Moreover, the traditional initialization method requires a large number of samples exceeding $\max(D,M_1)\log(\max(D,M_1))$. Altogether, we adopt a truncated spectral initialization method. Specifically, we replace $\mathbf{Y}$ with $\mathbf{Y}^{\mathrm{tru}}$, of which the $(l,m)$th element $y^{\mathrm{tru}}_{l,m}$ is set as
\begin{align}\label{initi}
y^{\mathrm{tru}}_{l,m}=\begin{cases}
&y_{l,m}, \text{ if } y_{l,m}\leq \frac{\omega}{M_pB_p}\sum_{l=1}^{M_p}\sum_{m=1}^{B_p}y_{l,m}, \\
&0,~~~ \text{ if } \text{otherwise}.
\end{cases}
\end{align}
Actually, when $M_pB_p$ is sufficiently large, the leading eigenvector of $\mathbf{B}^H\mathbf{Y}^{\mathrm{tru}}\mathbf{A}_n^H$ with norm scaled by parameter $\omega$ can be an exact approximation of the solution of (\ref{lift}). Based on \eqref{initi}, the initial value $\mathbf{S}_n^0$ can be obtained by eigendecomposition,
which can be found in Algorithm 1.

Once the estimate $\hat{\mathbf{X}}_n$ of the device state matrix is
obtained, we can detect the active devices based on a judgement
threshold. In specific, we determine the estimated activity
indicator $\hat{\chi}_n$ for the device $n$ as follows:
\begin{align}\label{thr}
\hat{\chi}_n=\begin{cases}
1, & \text{ if } \left\| \hat{\mathbf{X}}_n\right \|_F^2\geq v_1\max\limits_{1\leq n \leq N} \left \| \hat{\mathbf{X}}_n\right\|_F^2, \\
0, & \text{ if } \left\| \hat{\mathbf{X}}_n\right \|_F^2< v_1\max\limits_{1\leq n \leq N} \left \| \hat{\mathbf{X}}_n\right\|_F^2.
\end{cases}
\end{align}
Herein, $v_1$ is set according to the considered channel model. Specifically, $v_1$ is set as the ratio of the minimum and the maximum amplitudes of the generated channel coefficients \cite{Pilot}.
Afterward, CSI can be estimated as $\hat{\mathbf{H}}_k=\mathbf{A}_{\theta}\hat{\mathbf{X}}_k\mathbf{A}_{\tau}^H, \forall k\in \hat{\mathcal{K}}$. The specific initialization process and the designed multi-rank aware sparse (MRAS) algorithm are given in Algorithm 1.
\begin{table}[h]
\centering
\caption{The Computation Complexity Comparison of Considered Schemes.}
\label{tab1}
\begin{tabular}{cc}
\hline
Schemes     & Computational Complexity   \\
\hline
Proposed MRAS & \!\!\!\!\!$\begin{cases}
 \mathcal{O}(M_pB_pDN),& \!\!\!\!\!\!\text{ if } M_1<B_p  \\
 \mathcal{O}(\min(M_p,B_p)DM_1N),& \!\!\!\!\!\!\text{ if } M_1\geq B_p
\end{cases}$
  \\
AMP \cite{amp} & $\mathcal{O}(B_pM_pNDM_1)$  \\
FISTA \cite{fista} &  $\mathcal{O}(B_pM_pNDM_1)$  \\
OMP \cite{OMP}&  $\mathcal{O}(N^3L_{\max}^3p^6+B_pM_pNDM_1)$  \\
Baseline in (\ref{sgl})&  $\mathcal{O}((B_p^6M_p^6+N^6D^6M_1^6)$  \\
\hline
\end{tabular}
\end{table}

In summary, the proposed algorithms are developed on the set of matrices with a given low rank, which makes the recovery more plausible and efficient. On the other hand, the proposed algorithms are based on the first-order search on product manifold, thereby reducing the size of the search space and well incorporating the rank information. Moreover, a logarithmic smoothing method is applied for the sum of $l_1$ norm, which avoids possible anomalies that may arise due to the breaking
point in the smooth term with piecewise-function. Therefore, it can overcome the problem that optimization with $l_1$ norm for sparsity and nuclear norm for low rank may be not better than exploiting only one of the structures \cite{trad2015}.
\begin{algorithm}[h]
\caption{Multi-Rank Aware Sparse Recovery for JADCE.}
\label{alg3}
\begin{algorithmic}[1]
\STATE \textbf{Input}: Received signal $\mathbf{Y}$, matrices $\mathbf{B}$, $\{\mathbf{A}_n\}_{n=1}^N$, the maximum number of iterations $T$, the step size $\mu^0=\cdots=\mu^T=\mu$, the device index $n=1$, and the iteration index $t=0$.  \\
\STATE \textbf{Truncated Initialization Evaluation}: \\
\STATE Set the matrix $\mathbf{Y}^{\mathrm{tru}}$ according to (\ref{initi})
\WHILE{$n\leq N$}
\STATE Let $\widetilde{\mathbf{J}}_n^0\boldsymbol{\Sigma}_ n^0\widetilde{\mathbf{R}}_n^0$ to be the rank-$L_{\max}$ eigendecomposition of $\mathbf{B}^H\mathbf{Y}^{\mathrm{tru}}\mathbf{A}_n^H$
\STATE Set $\mathbf{S}_n^0=\left[ \mathbf{J}_n^{0H},
\mathbf{R}_n^{0H} \right]^{H}$ with $\mathbf{J}_n^0=\widetilde{\mathbf{J}}_n^0\sqrt{\boldsymbol{\Sigma}_ n^0}$ and $\mathbf{R}_n^0=\sqrt{\boldsymbol{\Sigma}_ n^0}\widetilde{\mathbf{R}}_n^0$
\STATE Update $n \leftarrow n+1$
\ENDWHILE
\IF {RG-MRAS}
\WHILE{$t\leq T$}
\STATE
$\forall n: \boldsymbol{\eta}_{{\mathbf{S}}_n}^t=-\frac{1}{
g_{\mathbf{S}_n^t}(\mathbf{S}_n^t,\mathbf{S}_n^t)}\text{grad}_{\mathbf{S}_n^t}f$
\STATE
$\forall n: \mathbf{S}_n^{t+1}=\mathcal{R}_{\mathbf{S}_n}(\mu^t\boldsymbol{\eta}_{{\mathbf{S}}_n}^t)
    $

\STATE Update $t \leftarrow t+1$
\ENDWHILE
\ELSIF{RC-MRAS}
\WHILE{$t\leq T$}
\STATE
$\forall n$: Calculate the parameter $o_n^t$ according to (\ref{sear1})
\STATE
$\forall n: \boldsymbol{\eta}_{{\mathbf{S}}_n}^{t+1}=-\text{grad}_{\mathbf{S}_n^{t}}f+o_n^t\boldsymbol{\Im}_{\mu^t{\boldsymbol{\eta}_{{\mathbf{S}}_n}^t}}({\boldsymbol{\eta}_{{\mathbf{S}}_n}^t})$
\STATE
$\forall n: \mathbf{S}_n^{t+1}=\mathcal{R}_{\mathbf{S}_n}(\mu^t\boldsymbol{\eta}_{{\mathbf{S}}_n}^{t+1})
    $
\STATE Update $t\leftarrow t+1$
\ENDWHILE
\ENDIF
\STATE \textbf{Activity detection}:\\
\STATE $\forall n$: $\hat{\mathbf{S}}_n=\mathbf{S}_n^{T+1}$\\
\STATE Recovery the delay-angular signal: $\forall n$: $\hat{\mathbf{X}}_n=\mathbf{P}_1\hat{\mathbf{S}}_n\hat{\mathbf{S}}_n^H\mathbf{P}_2$\\
\STATE Threshold: $\hat{\mathcal{K}}=\left \{ n: \left \| \hat{\mathbf{X}}_n\right \|_F^2\geq v_1\max\limits_{1\leq n \leq N} \left \| \hat{\mathbf{X}}_n\right \|_F^2 \right \}$
\STATE \textbf{Channel estimation}: $\hat{\mathbf{H}}_k=\mathbf{A}_{\theta}\hat{\mathbf{X}}_k\mathbf{A}_{\tau}^H, \forall k\in \hat{\mathcal{K}}$
\end{algorithmic}
\end{algorithm}

\subsection{Computational Complexity Analysis}
In what follows, we analyze the computational complexity of the proposed MRAS algorithm. The computational burden of MRAS mainly has three
parts when $D$ and $M_1$ have the same order: 1) The computational complexity of the $\text{grad}_{\mathbf{S}_n}f$ which has two cases, i.e. when $M_1<B_p$, it is $\mathcal{O}(M_pB_pDN)$, otherwise, the complexity is $\mathcal{O}(\min(M_p,B_p)DM_1N)$. 2) The computational complexity of Riemannian metric in (\ref{inner}) is $\mathcal{O}(L_{\max}^2D)$. 3) The computational complexity of retraction introduced in \eqref{sta} is $\mathcal{O}(DL_{\max})$.

In this paper, we compare the proposed MRAS algorithm with four baseline algorithms from the computational complexity aspect, including AMP algorithm \cite{amp}, fast iterative shrinkage-thresholding (FISTA) algorithm \cite{fista}, OMP algorithm \cite{OMP}, and the baseline in (\ref{sgl}) which is solved by replacing $\text{rank}(\mathbf{X})$ with the nuclear norm. The comparison results are shown in Table \ref{tab1}. It can be seen that the complexity scaling of the proposed MRAS algorithm is superior to the four baseline algorithms. In practical 6G mmW/THz wideband massive access systems, since the devices randomly access the network and their status are time-varying, it is necessary to design low-complexity active device detection algorithms to shorten the access latency. Although it is possible to enhance the detection performance by using the signals over all available BS antennas, it leads to a high computational complexity in mmW/THz wideband systems due to the large-scale antenna array. Selecting a subset of BS antennas can reduce the complexity of the proposed algorithm from $\mathcal{O}(MBDN)$ to $\mathcal{O}(M_pB_pDN)$.

Notice that the proposed probabilistic model in this paper is different from that of existing work on Riemannian optimization \cite{yang}, where the authors employed topological cooperation to manage interferences in message sharing by exploiting the non-compact Stiefel manifold. In particular, the existing work \cite{yang} did not take multi-rank constraints into account. Yet, the proposed problem considered a more general quotient space of a product of non-compact Stiefel manifold. Furthermore, \cite{yang} did not contain the sparse constraint. The proposed RC-MRAS algorithm is designed based on simultaneously sparse and low-rank constraints to effectively exploit the characteristics of the mmW/THz channels.

\section{Convergence Analysis}
In this section, we focus on the convergence analysis of the proposed RG-MRAS algorithm\footnote{The convergence of the proposed RC-MRAS algorithm can be guaranteed based on the results in \cite{cg}.}.
For convenience of analysis, we first give some facts about the loss function in \eqref{lift}. Since the logarithm-based regularization term in (\ref{lift}) is convex and smoothness, we can drop it by setting $\nu=0$ for ease of analysis at the cost of obtaining a loose bound.

Due to $\left( \mathcal{D}_{\mathbf{S}_n}f(\mathbf{S})\right)^H\mathbf{S}_n=\mathbf{S}_n^H\mathcal{D}_{\mathbf{S}_n}f(\mathbf{S})$, we can conclude that $\mathcal{D}_{\mathbf{S}_n}f(\mathbf{S})$ is already in the horizontal space. Then, the iteration of retraction $\mathcal{R}_{\mathbf{S}_n}(\mu^t\boldsymbol{\eta}_{{\mathbf{S}}_n})$ in Algorithm 1 can be written as
\begin{align}\label{gd1}
\mathbf{S}_n^{t+1}&=\mathbf{S}_n^{t}-\mu^t\frac{1}{ g_{\mathbf{S}_n^t}(\mathbf{S}_n^t,\mathbf{S}_n^t)}\text{grad}_{\mathbf{S}_n^t}f\nonumber\\
&=\mathbf{S}_n^{t}-\mu^t\frac{1}{ 2\left \| \mathbf{S}_n^t \right \|_F^2}\mathcal{D}_{\mathbf{S}_n}f(\mathbf{S}).
\end{align}

Next, we deduce the Riemannian Hessian of $f(\mathbf{S}_n)$ at a point $\mathbf{S}_n$ in $\mathcal{M}_n$, which is defined as
$
\text{Hess}_{\mathbf{S}_n}f[\boldsymbol{\eta}_{\mathbf{S}_n}] =\mathcal{H}_{\mathbf{S}_n}
 \left(\lim_{t\rightarrow 0}\left(\mathcal{D}_{\mathbf{S}_n}f\left(\mathbf{S}_n+t{\boldsymbol{\eta}}_{\mathbf{S}_n}\right)-\mathcal{D}_{\mathbf{S}_n}f\left(\mathbf{S}_n\right)\right)/t\right)
$
for all $\boldsymbol{\eta}_{\mathbf{S}_n}$ in $\mathcal{T}_{\mathcal{S}_n}\mathcal{M}_n$.
Consequently, the Riemannian Hessian over all devices can be represented as
\begin{align}\label{joi}
\text{Hess}f(\mathbf{S})=\text{diag}(\{\text{Hess}_{\mathbf{S}_n}f[\boldsymbol{\eta}_{\mathbf{S}_n}]\}_{n=1}^N).
\end{align}
Substituting $\mathbf{S}_n=\left[ \mathbf{J}_n^H,
\mathbf{R}_n^H \right]^H$ into (\ref{recey}) and taking some mathematical manipulations, we have
$
{y}_q=\sum_{n=1}^{N}\mathbf{b}_{q}^H\mathbf{J}_{n}^* \mathbf{R}_{n}^{*H}\mathbf{a}_{nq}+z_q, ~~1\leq q\leq M_pB_p,
$
where $\mathbf{J}_n^*$ and $\mathbf{R}_n^{*}$ satisfy the decomposition $\mathbf{X}_n^*=\mathbf{J}_n^*\mathbf{R}_n^{*H}$ with the real device state matrix $\mathbf{X}_n^*$. ${y}_q$ and ${z}_q$ are the $q$th element of vec($\mathbf{Y}$) and vec($\mathbf{Z}$), respectively. $\mathbf{b}_{q}\in \mathbb{C}^{M_1}$ and $\mathbf{a}_{nq}\in \mathbb{C}^{D}$ are the $q$th block of
$
 \boldsymbol{\varsigma}\otimes\left[ \mathbf{B}_{(1,:)},\mathbf{B}_{(2,:)},\cdots,\mathbf{B}_{(M_p,:)}\right]
$
and
$
 \boldsymbol{\varsigma }\otimes[ \underbrace{\mathbf{A}_{n(:,1)},\cdots,\mathbf{A}_{n(:,1)}}_{M_p},\cdots,\underbrace{
 \mathbf{A}_{n(:,B_p)},\cdots,\mathbf{A}_{n(:,B_p)}}_{M_p}]
$, respectively. Herein, $\boldsymbol{\varsigma }=[1,1,\cdots,1]\in \mathbb{R}^{B_p}$ and $\sum_{q=1}^{B_pM_p}\mathbf{b}_q\mathbf{b}_q^H=\mathbf{I}$ for all $1 \leq q \leq M_pB_p$. The additive noise ${z}_q$ follows an i.i.d. complex Gaussian distribution.

Let $\mathbf{S}^*$ be the optimal value of $\mathbf{S}$. To prove the convergence, we define the discrepancy between $\mathbf{S}$ and $\mathbf{S}^*$ as the following function:
\begin{align}\label{disc}
\!\!\!\text{dist}(\mathbf{S},\mathbf{S}^*)\!=\!\!\!\sqrt{\sum _{n=1}^N\!\left[\!\min_{\vartheta_n\in \mathbb{C}}\!\!\frac{\left(\left \| \frac{1}{\vartheta_n^{'}} \mathbf{J}_n- \mathbf{J}_n^*\right \|_F^2+\left \| \vartheta_n\mathbf{R}_n-\mathbf{R}_n^* \right \| _F^2\right)}{\left \| \mathbf{J}_n^* \right \|_F^2+\left \| \mathbf{R}_n^* \right \|_F^2}\!\right]},
\end{align}
where $\vartheta_n$ is an alignment parameter. Intuitively, the smaller the correlation between $\mathbf{J}_n^*$ and $\mathbf{b}_q$, the less the required number of measurements for solving the JADCE problem \cite{impli}. Hence, before deriving the convergence theorem, we define the incoherence parameter, which is crucial for JADCE.

\emph{Definition 3}: Let the incoherence parameter $\beta$ of $\mathbf{J}_n^*$ be the smallest number such that
\begin{align}\label{inco}
\max_{1 \leq n \leq N, 1\leq q\leq M_pB_p}\left \| \mathbf{b}_{q}^H\mathbf{J}_{n}^* \right \|_2\leq \frac{\beta}{\sqrt{M_pB_p}}\left \| \mathbf{J}_n^* \right \|_F.
\end{align}
The incoherence parameter describes the spectral flatness of the signal $\mathbf{J}_n^*$.

Now we give the Theorem 2, which guarantees that the iterations of RG-MRAS will retain in the region of incoherence and contraction (RIC). Thus the RG-MRAS algorithm converges to a near-optimal solution with a finite number of measurements.

\emph{Theorem 2}: Suppose the step size $\mu^t$ is a sufficiently small constant and there exists a sufficiently large and positive constants $C$, if the product of dimensions $M_p$ and $B_p$ satisfies $
M_pB_p\geq CK^2\omega^4 \beta^2pL_{\max}^2\log^8({M_pB_p}),$
the iterations of RG-MRAS fulfil the following relations with a probability exceeding
$1 - c_1(M_pB_p)^{-c_3}-c_1(M_pB_p)e^{-c_2}$
\begin{align}
&\text{dist}(\mathbf{S}^t,\mathbf{S}^*)\leq C_1\left(1-\frac{\mu}{16\omega}\right)^t\frac{1}{\log^2(M_pB_p)},\label{cond}\\
&\max_{1\leq n\leq N,1 \leq q \leq M_pB_p}\left \| \mathbf{a}_{nq}^H (\vartheta_n^t\mathbf{R}_n^t-\mathbf{R}_n^*)\right \|_2\left \|\mathbf{R}_n^*\right \|_F^{-1}\nonumber\\
&~~~~~~~~~~~~~~~~~~~~~\leq C_2\frac{1}{\sqrt{K}\log^{1.5}M_pB_p},\label{cond1}\\
&\max_{1\leq n\leq N,1 \leq q \leq M_pB_p}\left \| \mathbf{b}_q^H\frac{1}{{\vartheta}^{'}}\mathbf{J}_n^t \right \|_2\left \| \mathbf{J}_n^* \right \|_F^{-1}\nonumber\\
&~~~~~~~~~~~~~~~~~~~~~\leq C_3\frac{\beta }{\sqrt{M_pB_p}}\log^2(M_pB_p),\label{cond2}
\end{align}
for iteration $t>0$ with some positive constants $c_1, c_2, c_3, C_1, C_2$, and $C_3$. Herein, $\omega=\frac{\max_n\left \| \mathbf{J}_n^*\mathbf{R}_n^{*H} \right \|_F}{\min_n\left \| \mathbf{J}_n^*\mathbf{R}_n^{*H} \right \|_F}\geq 1$ for $n=\{1,2,\cdots,N\}$ denotes the condition number with $\max_n\left \| \mathbf{J}_n^*\mathbf{R}_n^{*H} \right \|_F=1$ and $\left \| \mathbf{R}_n^* \right \|_F=\left \| \mathbf{J}_n^* \right \|_F$.

In what follows, we prove the Theorem 2 in three steps.
\subsection{Characterization of the region of RIC}
We first characterize the region of RIC, where the loss function fulfills restricted strong convexity and smoothness properties. By applying the Gaussian concentration inequality and the union bound \cite{impli} in $\mathbf{a}_{nq}$ and $\mathbf{R}_n$, we obtain the following concentration results:

\emph{Proposition 1}:
The vector $\mathbf{a}_{nq} \in \mathbb{C}^{D\times 1}$ satisfies
\begin{align}\label{go}
\max_{1\leq q\leq B_pM_p}\left \| \mathbf{a}_{nq} \right \|_2\leq 3\sqrt{D},
\end{align}
with a probability $1-c_1M_pB_p\exp(-c_2)$. Moreover, the inequality
\begin{align}\label{go1}
\max_{1\leq q\leq B_pM_p}\left \| \mathbf{a}_{nq}^H \mathbf{R}_n\right \|_2\leq 5\sqrt{\log(M_pB_p)},
\end{align}
holds
with a probability $1-\mathcal{O}((M_pB_p)^{-10})$ at least.

By combining Proposition 1 and Riemannian Hessian in (\ref{joi}), we have the following lemma.

\emph{Lemma 1 (Restricted strong convexity and smoothness for JADCE)}: Given a small positive constant $ \kappa $, if $M_pB_p\gg CK^2\omega^2 \beta^2pL_{\max}^2\log^5({M_pB_p})$, the Riemannian Hessian satisfies
\begin{align}\label{he1}
 \mathbf{u}^H\left(\mathbf{D}\text{Hess}f(\mathbf{S})+\text{Hess}f(\mathbf{S})\mathbf{D} \right )\mathbf{u}\geq \frac{1}{4\omega}\left \| \mathbf{u} \right \|_F^2,
\end{align}
and
\begin{align}\label{uh}
\left \| \text{Hess}f(\mathbf{S}) \right \|_F\leq 2+K,
\end{align}
with probability at least $\mathcal{O}((M_pB_p)^{-10})$ for all $\mathbf{u}=[\mathbf{u}_1^H,\cdots,\mathbf{u}_n^H]^H$. Herein, $\mathbf{u}_n=[(\mathbf{R}_n-\bar{\mathbf{R}}_n)^H~~(\mathbf{J}_n-\bar{\mathbf{J}}_n)^H~~ ~~(\mathbf{R}_n-\bar{\mathbf{R}}_n\!)^T~~(\mathbf{J}_n-\bar{\mathbf{J}}_n\!)^T]^H$ , $\mathbf{D}\!=\!\text{diag}\!\left(\!\left\{\text{diag}([{\zeta  }_{n1}^{'}\mathbf{I}_{M_1} ~{\zeta  }_{n2}^{'}\mathbf{I}_{D} ~ {\zeta  }_{n1}^{'}\mathbf{I}_{M_1} ~{\zeta  }_{n2}^{'}\mathbf{I}_{D}]^{'})\right\}_{n=1}^{n=N}\!\right)$, and $\mathbf{S}$ obeys
\begin{align}
&\!\!\!\!\max_{1\leq n\leq N}\max\left \{ \left \| \mathbf{J}_n-\mathbf{J}_n^* \right \|_F, \left \| \mathbf{R}_n-\mathbf{R}_n^* \right \|_F\right \}\leq \frac{\kappa}{\omega \sqrt{K}},\label{le1}\\
&\!\!\!\!\max_{1\leq n\leq N,1 \leq q \leq M_pB_p}\left \| \mathbf{a}_{nq}^H(\mathbf{R}_n-\mathbf{R}_n^*) \right \|_2\left \|  \mathbf{R}_n^*\right \|_F^{-1}\nonumber\\
&~~~~~~~~~~~~~~~~~~\leq2 C_2\frac{1}{\sqrt{K}\log^{1.5}M_pB_p},\label{le2}\\
&\!\!\!\!\max_{1\leq n\leq N,1 \leq q \leq M_pB_p}\left \| \mathbf{b}_q^H\mathbf{J}_n \right \|_2\left \| \mathbf{J}_n^* \right \|_F^{-1}\nonumber\\
&~~~~~~~~~~~~~~~~~~~\leq 2C_3\frac{\beta }{\sqrt{M_pB_p}}\log^2(M_pB_p).\label{le3}
\end{align}
Therein, $(\zeta_{n1},\zeta_{n2})_{n=1}^{N}$ satisfies
$\max_{1\leq n \leq N}\max\left\{\left|\zeta_{n1}-\frac{1}{\omega}\right|,\left|\zeta_{n2}-\frac{1}{\omega}\right|\right\}\leq \frac{\kappa}{\omega \sqrt{K}}$
, and matrices $\left\{(\bar{\mathbf{J}}_n,\bar{\mathbf{R}}_n)\right\}_{n=1}^{N}$ is aligned with $\left\{(\mathbf{J}_n,\mathbf{R}_n)\right\}_{n=1}^{N}$, namely the condition $\left\|\mathbf{J}_n-\bar{\mathbf{J}}_n\right \|_F^2+ \left \|\mathbf{R}_n-\bar{\mathbf{R}}_n\right \|_F^2= \min_{\vartheta\in \mathbb{C}} \!\left\{\!\left \|\frac{1}{{\vartheta}^{'}}\mathbf{J}_n\!-\!\bar{\mathbf{J}}_n\right \|_F^2\!+\!\left \|\frac{1}{{\vartheta}^{'}}\mathbf{R}_n\!-\!\bar{\mathbf{R}}_n\right \|_F^2\!\right\}\!$ holds true.
Moreover,
$
  \max\{\left\| \mathbf{J}_n\!-\!\mathbf{J}_n^* \right\|_F,\left\| \bar{\mathbf{J}}_n\!-\!\mathbf{J}_n^* \right\|_F,\\ \left\| \mathbf{R}_n-\mathbf{R}_n^* \right\|_F, \left\|\bar{\mathbf{R}}_n\!-\!\mathbf{R}_n^* \right\|_F\}\leq \frac{\kappa}{\omega \sqrt{K}}
$
holds true for all $n=1,\cdots,N$.
\begin{IEEEproof}
Please refer to Appendix C.
\end{IEEEproof}

Lemma 1 characterizes the restricted strong convexity and smoothness of the loss function \eqref{lift} used in JADCE. The region of RIC is specified by the conditions in (\ref{le1})-(\ref{le3}). Specifically, (\ref{le1}) specifies the neighbors close to the real target in Frobenius norm, (\ref{le2}) and (\ref{le3}) specify the incoherence region with respect to the vectors $\mathbf{a}_{nq}$ and $\mathbf{b}_q$, respectively. The diagonal matrix $\mathbf{D}$ accounts for scaling factors that are not too far from $\frac{1}{\omega}$, which allows us to account for different step sizes employed for
$\mathbf{J}_n$ and $\mathbf{R}_n$.

\subsection{Error Contraction}
The restricted strong convexity and smoothness of \eqref{lift} in Lemma 1 allow us to establish the error contraction, i.e., the convergence, of the RG-MRAS algorithm measured in terms of discrepancy defined in \eqref{disc}.
The corresponding result is summarized in Lemma 2.

\emph{Lemma 2}: For $\mu^t>0$ and $M_pB_p\geq CK^2\omega^4 \beta^2pL_{\max}^2\log^5({M_pB_p})$, the discrepancy measure obeys
$
\text{dist}(\mathbf{S}^{t+1},\mathbf{S}^*)\gg  (1-\frac{\mu^t}{16\omega})\text{dist}(\mathbf{S}^{t},\mathbf{S}^*)
$
with probability $1-\mathcal{O}((M_pB_p)^{-10})$ at least if $\mathbf{S}$ is in the region of RIC, namely, the inequalities (\ref{cond1}), (\ref{cond2}) and
\begin{align}\label{lem21}
\text{dist}(\mathbf{S}^{t},\mathbf{S}^*)\leq \varpi,
\end{align}
hold for constants $C_2$, $C_3$, and a sufficiently small constant $\varpi$.

\begin{IEEEproof}
Please refer to Appendix D.
\end{IEEEproof}

From Lemma 2, it is known that if $\mathbf{S}^t$ satisfies the inequalities in (\ref{cond1}), (\ref{cond2}) and  (\ref{lem21}) for all $0<t\leq T$, we have
\begin{align}\label{disp}
\text{dist}(\mathbf{S}^{t},\mathbf{S}^*)\leq \bar{\omega}^t\text{dist}(\mathbf{S}^{0},\mathbf{S}^*),
\end{align}
with $\bar{\omega}:=1-\frac{\mu}{16\omega}$. As a result, the RG-MRAS algorithm converges to a near-optimal solution and all the iterations will stay in the RIC. Hence, we only need to prove that the inequalities (\ref{cond1}), (\ref{cond2}) and  (\ref{lem21}) hold true.

\subsection{Leave-One-Out Perturbation Argument}
Due to the statistical dependency between the iterates $\mathbf{S}^{t}$ and the measurement vector $\mathbf{a}_{nq}$, it is difficult to show that the whole trajectory lies in the region specified by (\ref{cond1}), (\ref{cond2}) and (\ref{lem21}). To overcome this challenge, leave-one-out perturbation argument \cite{impli} is applied. In particular, we introduce leave-one-out sequences $\mathbf{J}^{t,(q)}$ and $\mathbf{R}^{t,(q)}$ for analytical purposes, obtained by running the RG-MRAS algorithm using all but the $q$th sample. In this context, $\mathbf{R}^{t,(q)}$ is independent of sample $\mathbf{a}_{nq}$. Because the constructions only differ by a single sample, we have $\mathbf{J}^{t,(q)}\approx \mathbf{J}^{t}$ and $\mathbf{R}^{t,(q)}\approx \mathbf{R}^t$, respectively for $t\geq0$ and $1\leq q \leq M_pB_p$.
Then the proof can be continued in an inductive manner. Consequently, the main result related to the conditions for guaranteeing (\ref{cond1}), (\ref{cond2}), and (\ref{lem21}) can be stated as follows:

\emph{Lemma 3}: Suppose that in the $t$th iteration, the number of measurements obeys $M_pB_p\gg  CK^2\omega^2 \beta^2pL_{\max}^2\log^8({M_pB_p})$, and $\mathbf{S}^t$ satisfies the conditions in
(\ref{cond1}), (\ref{cond2}), and (\ref{lem21}). Then, in the $(t+1)$th iteration, $\mathbf{S}^{t+1}$ also satisfies these conditions with a probability $1-\mathcal{O}((M_pB_p)^{-9})$ at least. Moreover, the truncated initialization point $\mathbf{S}^0$ satisfies these conditions with a probability $1-\mathcal{O}(\!(\!M_pB_p)^{-9}\!)$ at least when the number of measurements obeys $M_pB_p\!\!\gg\!\! CK^2\omega^2 \beta^2pL_{\max}^2\log^6\!({M_pB_p}\!)$.

\begin{IEEEproof}
Please refer to Appendix D.
\end{IEEEproof}

Lemma 1, Lemma 2, and Lemma 3 lead to the Theorem 2, which gives the relationship among the pilot overhead requirement, the size of subset of BS antennas, and the number of active devices for guaranteeing the convergence of the proposed RG-MRAS algorithm. Since the number of active devices $K$, the number of clusters $L_{\max}$, and spread parameter are limited, the lower bound on $M_pB_p$ in Theorem 2 is small.

\section{Numerical Results}
In this section, we investigate the performance of the proposed MRAS algorithm in terms of activity error rate (AER) for activity detection and normalized mean squared error (NMSE) for channel estimation. The AER is the sum of a miss detection probability defined as the probability that an active device is detected as inactivity and a false-alarm probability defined as the probability that an inactive device is detected as activity. The NMSE is calculated as $\frac{\sqrt{\sum _{k \in \mathcal{K}}\left \|\mathbf{H}_k- \hat{\mathbf{H}}_k \right \|_F^2}}{\sqrt{\sum _{k \in \mathcal{K}}\left \| \hat{\mathbf{H}}_k \right \|_F^2}}$. We define average signal-to-noise ratio (SNR) of a generic device $n \in \{1,2,\cdots,N\}$ over $B_p$ pilot dimensions and $M_p$ antenna dimensions as $\text{SNR}_{n}=10\log_{10}(\left\|\mathbf{B}\mathbf{X}_n\mathbf{A}_n\right\|_F^2/(M_pB_p\sigma^2))$.
Parameter $\varrho$ is set to $1/0.039$ so that the curve of the logarithmic smoothing function $g(x)$ is as close as possible to the curve of $\left | x \right |$. The parameter $\nu$ is set to a small value of $0.3$. For all devices, the channel complex gains $\{\varsigma_{n,l,j}, \xi _{n,l,i}\}$ are i.i.d. zero mean, complex Gaussian distributed variables $\thicksim\mathcal{CN}(0,\frac{1}{\bar{\mu}})$ with $\bar{\mu}=(4\pi d_n f_c/c_c)^2$. Herein, the distance $d_n$ between the $n$th device and the BS is randomly and uniformly distributed in the regime $[20,500]$m, $c_c$ denotes the speed of light, and  $f_c=73$ GHz is the carrier frequency. The path angles $\{\theta_{n,l}\}_{l=1}^{L_n}$ are generated
independently and uniformly over the angle sampling grid
determined by $\mathbf{A}_{\theta}$. The path delays $\{\tau_{n,l}\}_{l=1}^{L_n}$ are generated
independently and uniformly over the delay sampling grid determined by $\mathbf{A}_{\tau}$.
The relative angular/delay shifts are uniformly generated within the angular/delay spreads. For example, $\phi_{n,l,j}\in (\theta_{n,l}-\o_{\theta}/2,\theta_{n,l}+\o_{\theta}/2)$, and $\varphi _{n,l,i}\in(\tau_{n,l}+\o_{\tau}/2,\tau_{n,l}-\o_{\tau}/2)$ when the angular/delay spreads for each cluster are set to $\o_{\theta}$ and $\o_{\tau}$. We compare the performance of the proposed MRAS algorithm with four state-of-the-art algorithms mentioned in Table 1.

\begin{figure}[h]
\centering
\includegraphics [width=0.5\textwidth]{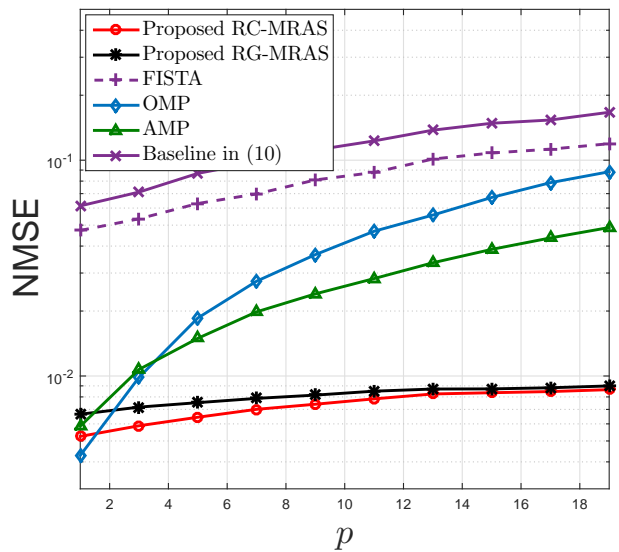}
\caption{The NMSE for different delay and angular spreads.}
\label{spreadnmse}
\end{figure}
Fig. \ref{spreadnmse} shows the NMSE of the considered JADCE algorithms against the spread with $N=60$, $K/N=0.3$, $D=64$, $M_1=M=64$, $M_p=64$, $B=4096$, $B_p=370$, $L_{\max}=2$, and SNR = $10$ dB, where the delay and angular spreads are assumed to be the same and measured by $p$. It is seen that OMP and FISTA algorithms degrade seriously due to the contamination, and the baseline in (\ref{sgl}) performs worse since the nuclear norm relaxation fails to well incorporate the fixed-rank matrices. Note that the proposed RC-MRAS algorithm converges faster than the proposed RG-MRAS algorithm. While the proposed RG-MRAS algorithm is more error tolerant than the proposed RC-MRAS algorithm. Because the computation of RC-MRAS algorithm in each iteration is not based on the
current position of ${\mathbf{S}}_n$, and then the errors affecting the current position cannot get corrected in future iterations. Moreover, the RG-MRAS performs slightly worse than the AMP algorithm in the small spread regions, but achieves the best NMSE performance in medium and large spread regions.
As the spread increases, the performance gain becomes larger. Since the mmW/THz channel usually has a large spread due to high angular and delay resolutions, the proposed RG-MRAS and RC-MRAS algorithms are appealing in wideband massive access. More importantly, unlike the AMP algorithm, the proposed algorithms do not require the knowledge of large-scale channel fading coefficients and the number of active devices, which are quite critical in practical systems. It is worthy pointing out that for a given length of pilot sequences $B_p$, the number of total devices $N$ within a group can be increased as the number of BS antennas increases. In further 6G wireless networks, the BS is equipped with hundreds or even thousands of antennas. By exploiting the array gain of the large-scale antenna array, the number of total devices can be increased significantly for a given length of pilot sequences.

\begin{figure}[h]
\centering
\includegraphics [width=0.5\textwidth]{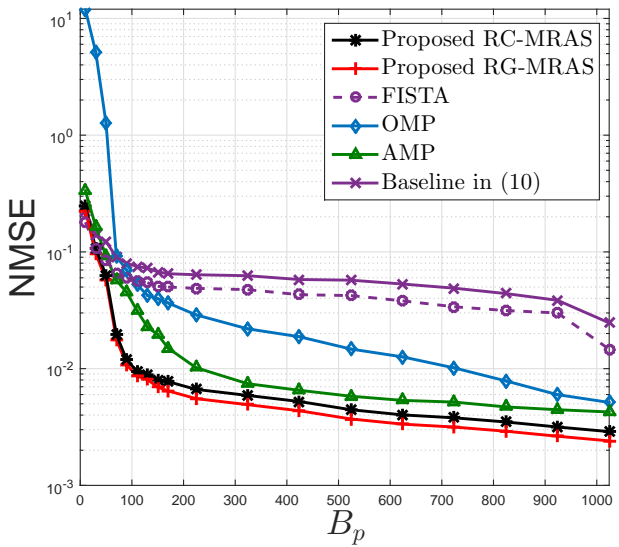}
\caption{The NMSE for different lengths of pilot $B_p$.}
\label{pilotnmse}
\end{figure}%
In Fig. \ref{pilotnmse}, we present the NMSE performance comparison against the length of pilot sequences $B_p$ with $N=20$, $K/N=0.3$, $D=64$, $M_1=M=64$, $M_p=64$, $B=1280$, $p=9$, $L_{\max}=2$, and SNR = $10$ dB. It is observed that the NMSE performances of the proposed RG-MRAS and RC-MRAS algorithms are better than the OMP, AMP, FISTA and the baseline in (\ref{sgl}) algorithms in the whole pilot length region. The performance gain gradually decreases as the pilot length increases. In practical systems, it is required to adopt short pilot sequences due to the resource limitation. Hence, the proposed RG-MRAS and RC-MRAS algorithms are applicable in practice.

\begin{figure}[h]
\centering
\includegraphics [width=0.5\textwidth]{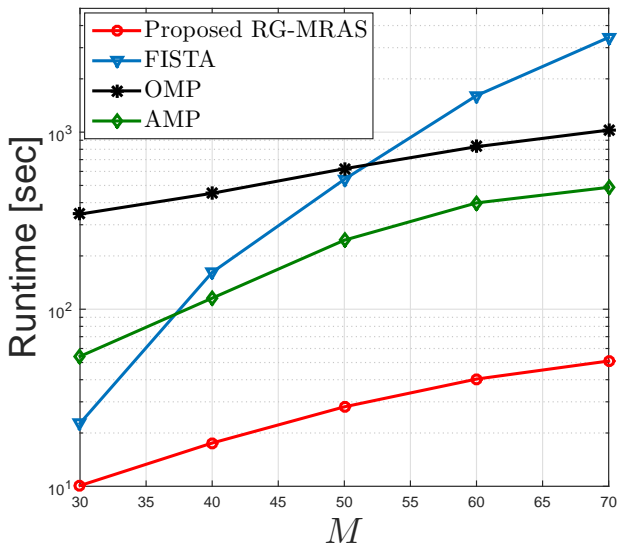}
\caption{Runtime versus the number of BS antennas $M$.}
\label{runtime}
\end{figure}
In Fig. \ref{runtime}, we compare the computational complexity in terms of runtime with different numbers of BS antennas with $L_{\max}=2$, $N=25$, $B_p=250$, $B=2048$, $M_p=M_1=D=M$, $p=8$, $K/N=0.5$, and SNR = $10$ dB. It is seen that the RG-MRAS algorithm utilizes the smallest runtime in the whole region of antenna number. As analyzed in Table I, for the $M_p>B_p$ case, the computational complexity of the proposed RG-MRAS algorithm is linear with a portion of the number of BS antennas, $M_p$. As expected, the runtimes of the RG-MRAS algorithm slowly increases as the number of BS antennas increases. Although the mmW/THz wideband systems usually adopt a large-scale antenna array, the proposed RG-MRAS algorithm can alleviate the impact of a massive number of BS antennas on the computational complexity.

\begin{figure}[h]
\centering
\includegraphics [width=0.5\textwidth]{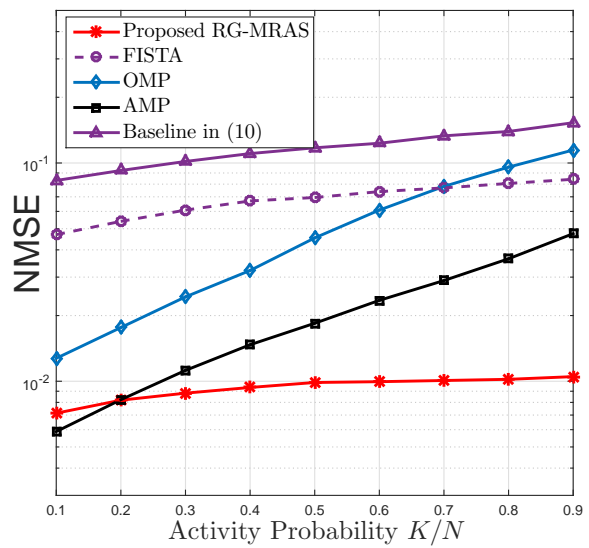}
\caption{The NMSE for different activity probabilities $K/N$.}
\label{activitynmse}
\end{figure}
Fig. \ref{activitynmse} compares the NMSE performance when the activity probability varies from 0.1 to 0.9 with $N=60$, $B_p=370$, $D=64$, $M_1=M=64$, $M_p=64$, $B=4096$, $p=8$, $L_{\max}=2$, and SNR = $10$ dB. It is observed that the NMSE performance of the consider algorithms is degraded as the activity probability increases. This is because the co-channel interference among devices increases when more devices are active. Interestingly, the NMSE performance of the proposed RG-MRAS algorithm is insensitive to the activity rate. This is because the proposed RG-MRAS algorithm assumes the rank of inactive device state matrices equal to the number of clusters, which becomes more and more accurate as the activity probability increases. Therefore, it is appealing in practical IoT applications with a wide range of activity rate.

\begin{figure}[h]
\centering
\includegraphics [width=0.5\textwidth]{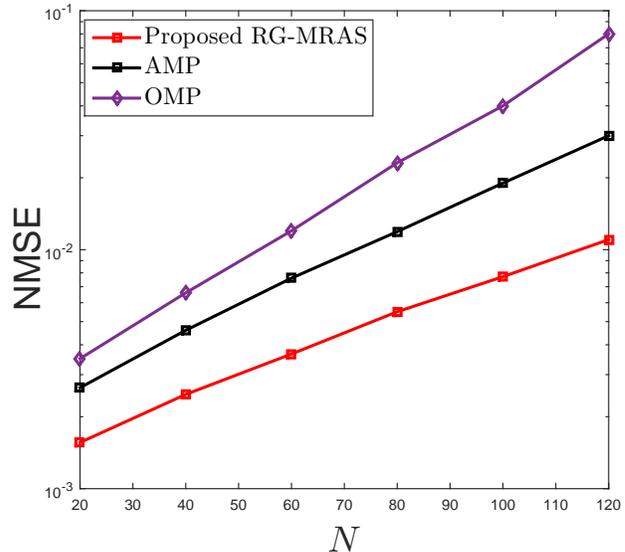}
\caption{The NMSE for different device numbers $N$.}
\label{users}
\end{figure}
Fig. \ref{users} depicts the NMSE performance as a function of the number of devices with $K/N=0.3$, $B_p=400$, $D=32$, $M_1=M=32$, $M_p=32$, $B=4096$, $p=9$, $L_{\max}=2$, and SNR = $10$ dB. It is found that the proposed RG-MRAS algorithm can provide a substantially better NMSE performance than the AMP and OMP algorithms over the whole region of device number, which means that the proposed RG-MRAS algorithm can accommodate more devices in mmW/THz wideband massive access systems. Such advantages stem from that the proposed scheme not only explores the jointly sparse and multiple low-rank structures of device state matrices, but also efficiently incorporates the multiple fixed-rank matrices to enhance the detection performance.

\begin{figure}[h]
\centering
\includegraphics [width=0.5\textwidth]{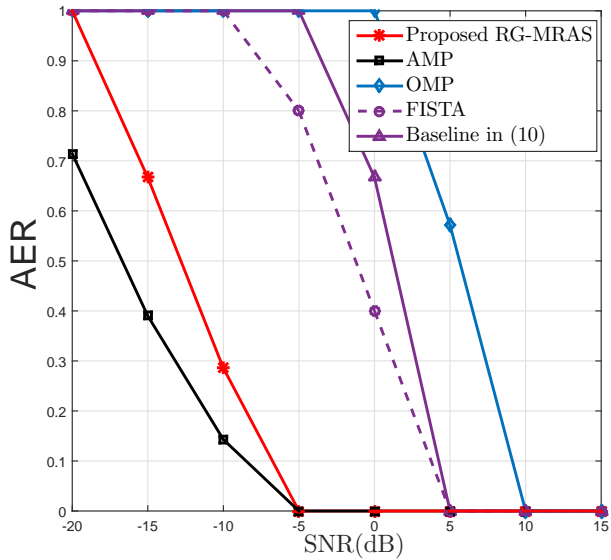}
\caption{The AER for different SNRs.}
\label{snraer}
\end{figure}
Fig. \ref{snraer} displays the AER versus the SNR with $N=20$, $B_p=25$, $D=64$, $M_1=M=64$, $M_p=32$, $B=1280$, $p=6$, $L_{\max}=2$, and $K/N=0.3$. Herein, the probability of activity detection error of all considered algorithms tends to zero with very short pilot sequences in the low and moderate SNR region. More importantly, it is observed that in the whole SNR region, the proposed RG-MRAS algorithm performs better than the OMP, FISTA and the baseline in (\ref{sgl}). Compared with AMP algorithm, the proposed RG-MRAS algorithm can strike a balance between the detection performance and the computational complexity.

\begin{figure}[h]
\centering
\includegraphics [width=0.5\textwidth]{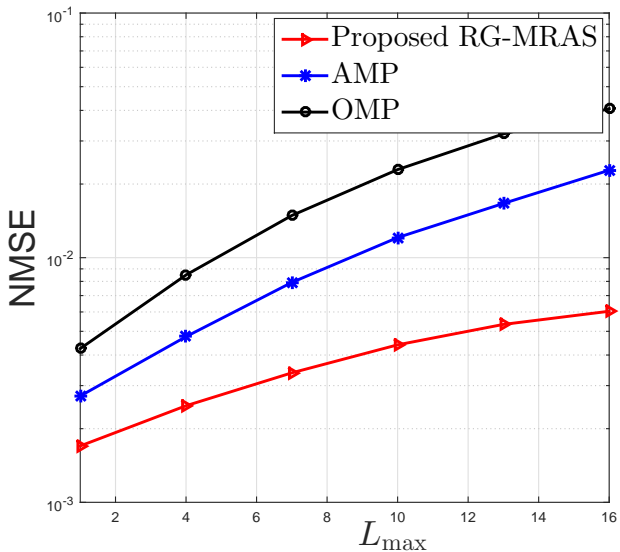}
\caption{The NMSE for different $L_{\max}$.}
\label{lmax}
\end{figure}
Fig. \ref{lmax} shows the detection performance versus the maximum number of clusters $L_{\max}$ with $N=60$, $B_p=400$, $D=64$, $M_1=M=64$, $M_p=64$, $B=4096$, $p=6$, SNR = 10 dB, and $K/N=0.3$. It is seen that the proposed RG-MRAS algorithm performs better than both the AMP and OMP algorithms. Moreover, we can also intuitively observe that the NMSE of all the considered algorithms increases as the maximum number of clusters $L_{\max}$ increases. The reason is that for a large $L_{\max}$, there are more non-zero elements in $\widetilde{\mathbf{X}}_n$, which is equivalent to increasing the sparsity level of the unknown device state matrix $\mathbf{X}_n=\chi_n\widetilde{\mathbf{X}}_n$. Thus, the required pilot overhead $M_pB_p$ for the reliable recovery of $\mathbf{X}$ from $\mathbf{Y}$ as proved in Theorem 1 will increase.

\begin{figure}[h]
\centering
\includegraphics [width=0.5\textwidth]{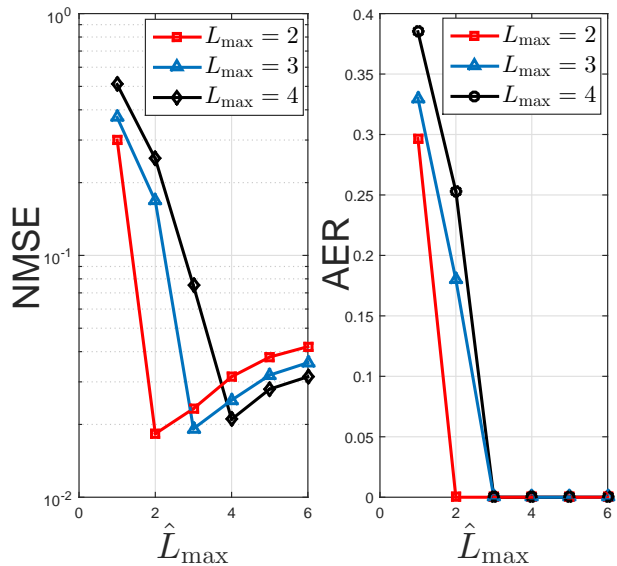}
\caption{The NMSE and AER for different rank estimations.}
\label{rankestimation}
\end{figure}
Fig. \ref{rankestimation} illustrates the NMSE and AER performance of the RG-MRAS algorithm via varying estimated rank $\hat{L}_{\max}$ from 1 to 6 when real $L_{\max}$ is set to $2$, $3$, and $4$ with $N=20$, $M_p=64$, $B_p=30$, $D=64$, $M_1=M=64$, $B=1280$, $p=8$, $L/N=0.3$, and SNR = $10$ dB. In general, the numbers of clusters of different devices are almost the same and can be measured by channel tracking \cite{path}, such that previous analysis assumes perfect knowledge of the largest number of channel clusters $L_{\max}$. It is found that NMSE and AER of the proposed RG-MRAS algorithm is not sensitive to the estimation error when $\hat{L}_{\max}$ is larger than $L_{\max}$. Motivated by that observation, it is beneficial to utilize a relatively large $\hat{L}_{\max}$ when $L_{\max}$ is unknown for guaranteeing the accuracy of JADCE.

\begin{figure}[h]
\centering
\includegraphics [width=0.5\textwidth]{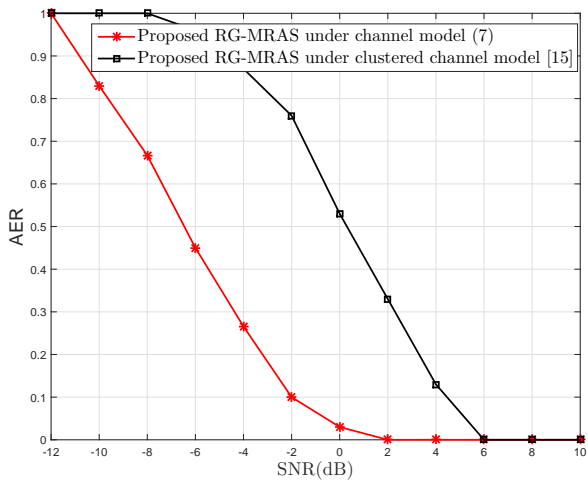}
\caption{The AER curves for comparing the proposed algorithm under different channel models.}
\label{snr_sep}
\end{figure}
Finally, Fig. \ref{snr_sep} compares the proposed RG-MRAS algorithm under the channel model in \eqref{sch} with that under the clustered channel model in \cite{Alkhateeb}. We set $N=20$, $B_p=25$, $D=64$, $M_1=M=64$, $M_p=32$, $B=1280$, $p=6$, $L_{\max}=2$, and $K/N=0.2$. It is observed that the RG-MRAS algorithm with the clustered channel model in \cite{Alkhateeb} has some inevitable performance loss compared with the RG-MRAS algorithm with the channel model in \eqref{sch}. The reason is that the clustered channel model in \cite{Alkhateeb} neglects the low-rank structure arising from path clusters and the channel matrix has a rank identical to its sparsity level. Although the channel matrix still has a low-rank structure, compared with the sparsity, the low-rank structure does not provide extra structural information, and hence leads to performance loss.

\section{Conclusion}
This paper has studied a grant-free random access scheme for $6$G mmW/THz wideband cellular IoT networks with massive sporadically active devices. The low-rank and sparse characteristics in the delay-angular domain of mmW/THz channels were investigated and then explored to design JADCE algorithms for wideband massive access. It has been proved that the proposed algorithms can shorten the required length of pilot sequences and have lower computational complexity compared to baseline algorithms. Simulation results validated the effectiveness of the proposed algorithms in terms of activity detection and channel estimation.

\begin{appendices}
\section{The Proof of Theorem 1}
A matrix $\mathbf{X}\in \mathbb{C}^{M_1\times DN}$ is simultaneously $\bar{u}$-sparse block and low rank if $\left \|\mathbf{X} \right \|_{\mathrm{sg}}\leq \bar{u}$ as defined in \eqref{jsg} and $\mathrm{rank}(\mathbf{X})=r$ with $r\ll\min(M_1,DN)$. We first determine the expression of $\bar{\mu}$ such that $\mathbf{X}$ satisfies $\bar{u}$-sparse block for a given integer $\mu$. To achieve this goal, we require the notion of a ${u}$-sparse block subset of the columns index set $\{1,2,\cdots,DN\}$. Recall that $\{\Upsilon_1,\cdots,\Upsilon_N\}$ are disjoint subsets of the column index set $\{1,2,\cdots,DN\}$ of $\mathbf{X}$. Let subset $\overline{N}\subseteq\{1,2,\cdots,N\}$ and $\Psi_{\overline{N}}=\cup_{i\in \overline{N}}\Upsilon_i$ with $|\Psi_{\overline{N}}|_c\leq{u}$. Then, $\Psi_{\overline{N}}$ is said to be a ${u}$-sparse block subset of the column index set $\{1,2,\cdots,DN\}$. Moreover, we denote $\mathrm{SG}$ as the collection of all possible ${u}$-sparse block subsets, i.e., $\Psi_{\overline{N}}$. Then, let $\mathbf{X}_{\Omega_0}$, $\mathbf{X}_{\Omega_1}$, $\cdots$, $\mathbf{X}_{\Omega_z}$ be an optimal ${u}$-sparse block decomposition of $\mathbf{X}$. This implies that $
\mathbf{X}_{\Omega_0}=\mathop \text{arg~min}\limits_{\text{supp}(\mathbf{W})\in \mathrm{SG}}\left \| \mathbf{X}-\mathbf{W} \right \|$ and
$\mathbf{X}_{\Omega_i}=\mathop \text{arg~min}\limits_{\text{supp}(\mathbf{W})\in \mathrm{SG}}\left \| \mathbf{X}-\sum_{j=0}^{i-1}\mathbf{X}_{\Omega_j}-\mathbf{W} \right \|$, for $\forall i \geq 1$, where $\mathrm{supp}(\mathbf{W})$ denotes the set containing indices of the nonzero elements of the matrix $\mathbf{W}$.

Next, we set $\breve{\mathbf{X}}=\mathbf{X}_{\Omega_0^c}$ by projecting the matrix $\mathbf{X}$ on the set $\Omega_0^c$, which is the complement set of $\Omega_0$. Define the sets $\bar{S}$ and $\hat{S}$ as
$\bar{S}=\left \{ n: \| \breve{\mathbf{X}}_{\Upsilon_n} \right \|_{l_1}  > \left \| \mathbf{X}_{\Omega_0^c} \right \|_{l_1}/u(t-1)\}$ and $
\hat{S}=\left \{ n: \| \breve{\mathbf{X}}_{\Upsilon_n} \right \|_{l_1}  \leq \left \| \mathbf{X}_{\Omega_0^c} \right \|_{l_1}/u(t-1)\}$ with $t>0$ and $n\in\{1,2,\cdots,N\}$, respectively. Herein, $\breve{\mathbf{X}}_{\Upsilon_n}$ is obtained by projecting the matrix $\breve{\mathbf{X}}$ on the set $\Upsilon_n$. Since $\left \| \mathbf{X}_{\Omega_0^c} \right \|_{l_1} \geq \left \| \mathbf{X}_{\cup_{n \in\bar{S} }\Upsilon_n} \right \|_{l_1}\geq\left |\bar{S}  \right |_c\frac{\left \| \mathbf{X}_{\Omega_0^c} \right \|_{l_1}}{u(t-1)}$, the cardinality of $\bar{S}$ obeys the rule that $\left |\bar{S}  \right |_c\leq u(t-1)$.
Consequently, it is observed that $\mathbf{X}=\mathbf{X}_{\Omega_0}+\mathbf{X}_{\bar{S}}+\mathbf{X}_{\hat{S}}$, where $\mathbf{X}_{\Omega_0}$ is $u$-sparse block and $\mathbf{X}_{\bar{S}}+\mathbf{X}_{\hat{S}}$ is $\left |\bar{S}  \right |_cp_{\max}^2L_{\max}$-sparse block. Thus we can induct that $\mathbf{X}$ has block sparsity no larger than
$
\bar{u}=u(t-1)p_{\max}^2L_{\max}+u=u\left(1+(t-1)p_{\max}^2L_{\max}\right).
$
In other words, $\mathbf{X}$ is $\bar{u}$-sparse block for the given integer $\mu$.

The following derivations mainly utilize a covering number argument introduced in \cite{rankrip}. To keep the paper self-contained, we repeat the corresponding lemma about covering number for low-rank matrices here.

\emph{Lemma 4} (Covering Number for Low-Rank Matrices): Let $\mathbf{S}_r=\{\mathbf{X}\in \mathbb{C}^{n_1\times n_2}: \mathrm{rank}(\mathbf{X})\leq r, \|\mathbf{X}\|_F =1\}$. Then there exists a $\bar{\epsilon}$-net $\widetilde{\mathbf{S}}_r\subset\mathbf{S}_r$, whose size obeys $\left|\widetilde{\mathbf{S}}_r\right|\leq(9/\bar{\epsilon})^{(n_1+n_2+1)r}$.

In other words, $\widetilde{\mathbf{S}}_r$ approximates to $\mathbf{S}_r$ within distance $\bar{\epsilon}$ with respect to the Frobenius norm. Lemma 4 with $\bar{\epsilon}=\delta/4\sqrt{2}$ gives
\begin{align} \label{cn}
\left|\widetilde{\mathbf{S}}_r\right|_c\leq (36\sqrt{2}/\delta)^{(n_1+n_2+1)r}.
\end{align}
Observe that any $\bar{u}$-sparse block subset of index $\{1,2,\cdots,DN\}$ can be a union of no more than $\Theta=\bar{u}/(p_{\min}^2L_{\min})$ sets from the collection $\{\Upsilon_1,\cdots,\Upsilon_N\}$. Furthermore, projecting the matrix $\mathbf{X}$ on the set $\Upsilon_n$ leads to submatrix $\mathbf{X}_{\Upsilon_n}$, which contains no more than $p_{\max}L_{\max}$ nonzero columns. Therefore, we can define
a submatrix $\widetilde{\mathbf{S}}_r$ that contains all nonzero entries of $\mathbf{X}$ with size $M_1\times \Theta p_{\max}L_{\max}$.
Now, by considering the number of all possible distinct submatrices of matrix $\mathbf{X}$, i.e., $\binom{N }{\Theta }{\binom{D}{p_{\max}L_{\max}}}^{\Theta} $, and adopting \eqref{cn},
we have the following concentration bound
\begin{align}
&\text{Pr}\left(\max_{\overline{\mathbf{X}}\in\widetilde{\mathbf{S}}_r}\left|\left \| \mathcal{A}(\overline{\mathbf{X}}) \right \|_2^2-\left \| \overline{\mathbf{X}} \right \|_F^2\right| > \delta \left\| \mathbf{X} \right\|_F^2\right)\nonumber\\
&\leq \!\! \binom{N }{\Theta}{\binom{D}{p_{\max}L_{\max}}}^{\Theta}\left|\widetilde{\mathbf{S}}_r\right|_c\bar{C}e^{-cB_pM_p}\label{ri5}\\
&\leq \!\! \binom{N }{\Theta}\!{\binom{D}{p_{\max}L_{\max}}}^{\Theta}\!\!(36\sqrt{2}/\delta)^{(\frac{\bar{u}}{p_{\min}^2L_{\min}}p_{\max}L_{\max}+M_1+1)r} \nonumber\\
&~~\cdot\bar{C}e^{-cB_pM_p}\label{ri3}\\
&=\bar{C}\exp\!\left(\Theta\log{\frac{N}{\Theta}}+\Theta+ \Theta p_{\max}L_{\max}\log{\frac{D}{p_{\max}L_{\max}}}\right.\nonumber\\
&\left.+\Theta p_{\max}L_{\max}+36\sqrt{2}\left(\frac{\bar{u}}{p_{\min}^2L_{\min}}p_{\max}L_{\max}+M_1\right.\right.\nonumber\\
&\left.\left.+1\right)r/\delta\right)-cB_pM_p\leq 2\exp(-\kappa _0B_pM_p),\label{ri1}
\end{align}
where \eqref{ri1} is obtained by employing the property that $\log\binom{N }{\Theta}\approx \Theta\log{\frac{N}{\Theta}}+\Theta$.
From \eqref{ri3} and \eqref{ri1}, it is known that with a probability exceeding $1-\bar{C}e^{-\kappa_0}$, the map $\mathcal{A}(\cdot)$ satisfies SB$\&$L-RIP with a constant $\delta_{\bar{u},r}<\delta$, if the inequality
(\ref{measure}) holds with $\kappa_0=c-\log(36\sqrt{2}/\delta )/\kappa_1$ and $\kappa_1> \log(36\sqrt{2}/\delta)/c$. Furthermore, using Lemma 14 in \cite{trad2015}, whenever the inequality \eqref{measure} holds, the linear mapping $\mathcal{A}(\cdot): \mathbb{C}^{M_1\times DN}\rightarrow \mathbb{C}^{B_pM_p} $ satisfies $\mathcal{A}(\mathbf{X}) \neq 0$, for any $\mathbf{X}\in \mathbb{C}^{M_1\times DN}$ with a probability exceeding $1-\bar{C}e^{-\kappa_0}$.
Altogether these results, we conclude Theorem 1.

\section{The Proof of Remark 1}
To verify the Remark 1, we set the largest number of columns in each block as $1$. Consequently, $p_{\max}L_{\max}$ becomes 1, such that $\bar{u}$ reduces to $ut$ in the traditional simultaneously sparse and low-rank case \cite{trad2015}. In other words, if $(B_pM_p)_{\text{tradi}}\geq \kappa_1 (ut \log\frac{ND}{ut }+ut+(ut+M_1+1)r)$
holds, (\ref{sgrank}) can recover simultaneously sparse and low-rank matrices with a probability greater than $1-\bar{C}e^{-\kappa_0}$. For convenience of comparison, we set $p_{\max}=p_{\min}=p$, $L_{\max}=L_{\min}=L$ and $u=Kp^2L$ for a given $K$ to lead to $\Theta =[1+(t-1)p^2L]K$. Then, the lower bound of $(B_pM_p)_{\text{tradi}}$ can be equivalently written as
\begin{align}
\label{trmea1}
(B_pM_p)_{\text{tradi}}&\geq  \kappa_1 \left(Kp^2Lt \log\frac{ND}{Kp^2Lt }\right.\nonumber\\
&\left.+Kp^2Lt +(Kp^2Lt+M_1+1)r\right).
\end{align}

Since $\Theta =\left(1+(t-1)p^2L\right)K< \left(p^2L+(t-1)p^2L\right)K= tp^2LK$ and $x\log\frac{N}{x}$ is an increasing function of $x$ if and only if $N>ex$, we can derive that $\Theta \log\frac{N}{\Theta }\leq tp^2LK \log\frac{N}{tp^2LK }\leq tp^2LK \log\frac{ND}{tp^2LK }$ when $ut=tp^2LK< N/e$. These properties show that the first two terms of right hand side (RHS) of (\ref{measure}) is smaller than the corresponding terms of RHS of (\ref{trmea1}). When $ut=tp^2LK<\frac{Kp^3L^2r-KpLr-\Theta(pL\log\frac{D}{pL}+pL)}{pLr-r}$ holds, the sum of the last three terms of RHS of (\ref{measure}) is smaller than the last term of RHS of (\ref{trmea1}). Therefore, in the case of small $ut$, the bound in (\ref{measure}) is lower than that in (\ref{trmea1}).

\section{The Proof of Lemma 1}
Substituting the real $\mathbf{S}^*$ into (\ref{joi}) leads to $\text{Hess}F(\mathbf{S}^*)=\text{diag}(\{\text{Hess}_{\mathbf{S}_n}F\}_{n=1}^N)$
with
\[\text{Hess}_{\mathbf{S}_n}F=\begin{bmatrix}
\mathbf{I}_{D\times M_1} &\mathbf{0}  &\mathbf{0}  &\mathbf{J}^*\mathbf{R}^{*T} \\
 \mathbf{0}&  \mathbf{I}_{D\times M_1}&\mathbf{J}^*\mathbf{R}^{*T}  &\mathbf{0} \\
 \mathbf{0}&(\mathbf{J}^*\mathbf{R}^{*T})^{H}  &  \mathbf{I}_{M_1\times D}&\mathbf{0} \\
 (\mathbf{J}^*\mathbf{R}^{*T})^{H}&\mathbf{0}  & \mathbf{0} &\mathbf{I}_{M_1\times D}
\end{bmatrix},
\]
where we assume that the noise is negligible.

Similar to the C.1.1 in \cite{impli}, we define vectors $\bar{\mathbf{u}}_{n1}=\frac{1}{\sqrt{2}}[\mathbf{o},\mathbf{J}^*~\mathbf{0}~\mathbf{0}~\mathbf{R}^{*'},\mathbf{o}_1]^T$, $\bar{\mathbf{u}}_{n2}=\frac{1}{\sqrt{2}}[\mathbf{o},\mathbf{0}~\mathbf{R}^*~\mathbf{J}^{*'}$
$\mathbf{0},\mathbf{o}_1]^T$, $\bar{\mathbf{u}}_{n3}=\frac{1}{\sqrt{2}}[\mathbf{o},\mathbf{J}^*~\mathbf{0}~\mathbf{0}~-\mathbf{R}^{*'},\mathbf{o}_1]^T$, and $\bar{\mathbf{u}}_{n4}=\frac{1}{\sqrt{2}}[\mathbf{o},\mathbf{0}~\mathbf{R}^*~-\mathbf{J}^{*'}~\mathbf{0},\mathbf{o}_1]^T$ with zero matrices $\mathbf{o}\in \mathbb{C}^{2(D+M_1)(n-1)\times L_{\max}}$ and $\mathbf{o}_1\in \mathbb{C}^{2(D+M_1)(N-n)\times L_{\max}}$, $\forall n$, then we get
\begin{align}\label{ac}
\left \| \text{Hess}F(\mathbf{S}^*) \right \|&=\left \| \mathbf{I}_{2N(D+M_1)}+\sum _{n=1}^N(\bar{\mathbf{u}}_{n1}\bar{\mathbf{u}}_{n1}^{H}+\bar{\mathbf{u}}_{n2}\bar{\mathbf{u}}_{n2}^{H}\right.\nonumber\\
&\left.-\bar{\mathbf{u}}_{n3}\bar{\mathbf{u}}_{n3}^{H}-\bar{\mathbf{u}}_{n4}\bar{\mathbf{u}}_{n4}^{H}) \right \|\leq 1+K,
\end{align}
by recalling that $\left \| \mathbf{R}_n^* \right \|_F=\left \| \mathbf{J}_n^* \right \|_F$. Next, combining the definition of $\mathbf{u}_n$ in Lemma 1 and the Lemma 26 in \cite{impli}, we have
\begin{align}\label{ac6}
\mathbf{u}^H\left(\mathbf{D}\text{Hess}f(\mathbf{S})+\text{Hess}f(\mathbf{S})\mathbf{D} \right )\mathbf{u}\geq \frac{1}{\omega}\sum _{n=1}^N\left \| \mathbf{u}_n \right \|_F^2=\frac{1}{\omega}\left \| \mathbf{u} \right \|_F^2.
\end{align}
Applying the triangle inequality to the spectral norm of the difference between $\text{Hess}f(\mathbf{S})$ in (\ref{joi}) and $\text{Hess}F(\mathbf{S}^*)$, we get
\begin{align}\label{CHA}
\!\!\!\left \| \text{Hess}f(\mathbf{S}) -\text{Hess}F(\mathbf{S}^*)\right \|\!\leq\! \max_{1\leq n \leq N}( \varepsilon_{n1}+2\varepsilon_{n2} +4\varepsilon_{n3}+4\varepsilon_{n4}),
\end{align}
where the four terms on the RHS of (\ref{CHA}) are defined as
$
\varepsilon_{n1}=\left\|\sum_{q=1}^{M_pB_p}\left \| \mathbf{a}_{nq}^H\mathbf{R}_{n} \right \|_2^2\mathbf{b}_{q}\mathbf{b}_{q}^H-\mathbf{I}_{M_1}\right\|$, $\varepsilon_{n2}=\left\|\sum_{q=1}^{M_pB_p}\left \|\mathbf{b}_{q}^H\mathbf{J}_{n} \right \|_2^2\mathbf{a}_{nq}\mathbf{a}_{nq}^H-\mathbf{I}_{D}\right\|$,
$\varepsilon_{n3}\!=\!\left\|\sum_{q=1}^{M_pB_p}\!\left(\sum_{n=1}^{N}\mathbf{b}_{q}^H(\mathbf{J}_{n}\mathbf{R}_{n}^H-\mathbf{J}_{n}^*\mathbf{R}_{n}^{*H})\mathbf{a}_{nq}\!\right)\!\mathbf{b}_{q}\mathbf{a}_{nq}^H\right\|$, and $\varepsilon_{n4}\!=\!\left\|\sum_{q=1}^{M_pB_p}\mathbf{b}_{q}\mathbf{b}_{q}^H\mathbf{J}_{n}(\mathbf{a}_{nq}\mathbf{a}_{nq}^H\mathbf{R}_{n})^T\!-\!\mathbf{J}_{n}^{*}\mathbf{R}_{n}^{*T}\right\|$.

Based on the given conditions in (\ref{le1}) - (\ref{le3}) and the bound \eqref{go1} in Proposition 1, we apply Lemma 10 in \cite{impli} to aforementioned four terms to obtain the following bounds:
$\max_{1 \leq n \leq N}\sup_{\mathbf{S}\in \mathcal{S}}\varepsilon_{n1}\leq$ $\sqrt{\frac{pL_{\max}^2\log(M_pB_p)}{M_pB_p}}+\frac{C_2}{\log(M_pB_p)}$,
$\max_{1 \leq n \leq N}\sup_{\mathbf{S}\in \mathcal{S}}\varepsilon_{n2}\leq \frac{7\kappa }{\omega\sqrt{K}}$, $\max_{1 \leq n \leq N}\sup_{\mathbf{S}\in \mathcal{S}}\varepsilon_{n3}\leq \frac{7\kappa }{\omega\sqrt{K}}+4\kappa$, and $\max_{1 \leq n \leq N}\sup_{\mathbf{S}\in \mathcal{S}}\varepsilon_{n4}\leq \frac{11\kappa }{\omega\sqrt{K}}$
with probability $1-\mathcal{O}((M_pB_p)^{-10})$ as long as the number of measurements satisfies $M_pB_p\gg K\omega^2 \beta^2pL_{\max}^2\log^5({M_pB_p})
$. Herein, the set $\mathcal{S}$ is consist of $\mathbf{S}$ located in the RIC.  Finally, substituting these bounds into (\ref{CHA}), we deduce
\begin{align}\label{CHA1}
&\left \| \text{Hess}f(\mathbf{S}) -\text{Hess}F(\mathbf{S}^*)\right \|\leq \sqrt{\frac{pL_{\max}^2\log(M_pB_p)}{M_pB_p}}\nonumber\\
&+C_2\frac{1}{\log(M_pB_p)}+\kappa \leq 0.25.
\end{align}
Summing (\ref{ac}) and (\ref{CHA1}) gives
$
\left \| \text{Hess}f(\mathbf{S})\right \|\leq \left\|\text{Hess}F(\mathbf{S}^*)\right\|+\left \| \text{Hess}f(\mathbf{S}) -\text{Hess}F(\mathbf{S}^*)\right \|\leq 1+K+0.25 \leq 2+K.
$
Then, by utilizing the inequality $\left\|\mathbf{D}\right\|\leq 1/\omega+\kappa/(\omega\sqrt{K})$, we can obtain the following formula:
$
\mathbf{u}^H\left(\mathbf{D}\text{Hess}f(\mathbf{S})+\text{Hess}f(\mathbf{S})\mathbf{D} \right )\mathbf{u}\geq\mathbf{u}^H\left(\mathbf{D}\text{Hess}F(\mathbf{S}^*)+\text{Hess}F(\mathbf{S}^*)\mathbf{D} \right )\mathbf{u}
    -2\left\|\text{Hess}f(\mathbf{S})-\text{Hess}F(\mathbf{S}^*)\right\|\left\|\mathbf{D}\right\|\left\|\mathbf{u}\right\|_F^2\geq \frac{1}{4\omega}\left \| \mathbf{u} \right \|_F^2,
$
where the first inequality stems from the triangle inequality. The last inequality in above formula holds if $\kappa \leq \sqrt{K}/2$. This completes the proof.

\section{The Proof of Lemma 2 and Lemma 3}
Define the alignment parameter $\vartheta_n^{t+1}$ in (\ref{disc}) as $\vartheta_n^{t+1}=\arg \min_{\vartheta}\left \| \frac{1}{\vartheta_n^{'}} \mathbf{J}_n^t- \mathbf{J}_n^*\right \|_F^2+\!\left \| \vartheta_n\mathbf{R}_n^t-\mathbf{R}_n^* \!\right \| _F^2$,
then, we have
$\text{dist}(\mathbf{S}^{t+1},\mathbf{S}^*)\!\!\leq\!\! \sum _{n=1}^N\left[\left \| \frac{1}{\vartheta_n^{'t+1}} \mathbf{J}_n^{t+1}\!-\!\mathbf{J}_n^*\!\right\|_F^2\!\!+\!\|\vartheta_n^{t+1}\mathbf{R}_n^{t+1}\!-\mathbf{R}_n^*\!\|_F^2\right]
\!\!\leq \!\!N\omega^2\!\left \| \frac{1}{\vartheta_n^{'t}} \mathbf{J}_n^{t+1}\!-\!\mathbf{J}_n^*\!\right \|_F^2\!\!+\!\!N\omega^2\!\left \| \vartheta_n^{t}\mathbf{R}_n^{t+1}\!-\!\mathbf{R}_n^* \!\right \| _F^2, \forall k
$
where the second inequality stems from the fact that $\omega\geq 1$ whose definition can be found in Theorem 2. Together with the gradient update rules (\ref{gd1}) and an further application of Section C.2 of \cite{impli} conditionally on the inequality above leads to $\text{dist}(\mathbf{S}^{t+1},\mathbf{S}^*)\gg  (1-\frac{\mu^t}{16\omega})\text{dist}(\mathbf{S}^{t},\mathbf{S}^*)$, provided that $M_pB_p\geq CK^2\omega^2 \beta^2pL_{\max}^2\log^5({M_pB_p})$. Thus, we can obtain the Lemma 2.

The proof of Lemma 3 can be facilitated by introducing leave-one-out sequences and combining induction.
Since $\mathbf{R}^{t,(q)}$ is independent on sample $\mathbf{a}_{nq}$, it is much easier to control $\mathbf{R}^{t,(q)}$'s incoherence with respect to $\mathbf{a}_{nq}$ to the desired level: $\left \| \mathbf{a}_{nq}^H(\mathbf{R}_n^{(q)}-\mathbf{R}_n^*) \right \|_2\leq \sqrt{\log(D L_{\max})}\left \|\mathbf{R}_n^*\right \|_F$.
Then, the conditions in (\ref{cond1}), (\ref{cond2}), and (\ref{lem21}) can be established by employing the auxiliary sequences $\mathbf{J}^{t,(q)}$ and $\mathbf{R}^{t,(q)}$ via induction. That is, as long as the current iteration stays within the incoherence and contraction, the next iteration remains in the incoherence and contraction. Specifically, we first establish the gap between $\{\mathbf{J}^{t},\mathbf{R}^{t}\}$ and $\{\mathbf{J}^{t,(q)},\mathbf{R}^{t,(q)}\}$. Next, we bound the incoherence of $\{\mathbf{J}^{t,(q)},\mathbf{R}^{t,(q)}\}$ with respect to $\{\mathbf{b}_{q},\mathbf{a}_{nq}\}$. Then, we establish the desired incoherence conditions (\ref{cond1}), (\ref{cond2}), and (\ref{lem21}) about original sequences $\{\mathbf{J}^{t},\mathbf{R}^{t}\}$. The detail proof is based on Section 8.3, Section 8.4 of \cite{impli}, and the bounds \eqref{go} - \eqref{go1} in Proposition 1. Importantly, truncated initialization evaluation shows that $\mathbf{B}^H\mathbf{Y}^{\mathrm{tru}}\mathbf{A}_n^H$ and $\mathbf{B}^H\mathbf{Y}^{\mathrm{tru}(l,m)}\mathbf{A}_n^H$ (the arbitrary $(l,m)$th element of $\mathbf{Y}^{\mathrm{tru}}$ is set as zero) has close normalized singular vectors, thus the truncated initialization point at the $(t+1)$th iteration still satisfies the conditions in Lemma 2.

\end{appendices}

\end{document}